\pdfoutput=1
\documentclass[epj]{svjour}
\usepackage{graphics}
\usepackage{graphicx}
\usepackage{atlasphysics}
\usepackage[hyperindex,breaklinks]{hyperref} 
\usepackage{url}
\newcommand{\dedx}{\ensuremath{\mathrm{d}E/\mathrm{d}x}}
\newcommand{\dEdx}{\ensuremath{\mathrm{d}E/\mathrm{d}x}}
\newcommand{\lumi}{18.4}
\usepackage{lineno}

\usepackage{preprintcover}

\PreprintCoverPaperTitle{Search for metastable heavy charged particles
with large ionisation energy loss in \emph{pp} collisions at \boldmath{$\sqrt{s} = 8$}~TeV  using the \mbox{ATLAS} experiment} 

\PreprintIdNumber{CERN-PH-EP-2015-124}  

\PreprintCoverAbstract{Many extensions of the Standard Model predict the existence of charged heavy long-lived particles, such as $R$-hadrons or charginos.
These particles, if  produced at the Large Hadron Collider, should be moving non-relativistically and are therefore identifiable through the measurement of an anomalously large specific energy loss in the \mbox{ATLAS} pixel detector. Measuring heavy long-lived particles through their track parameters in the vicinity of the interaction vertex provides sensitivity to metastable particles with lifetimes from 0.6~ns to 30~ns.
A search for such particles with the \mbox{ATLAS} detector at the Large Hadron Collider is presented, based on a data sample corresponding to an integrated luminosity of $\lumi$~fb$^{-1}$ of  $pp$ collisions at $\sqrt{s} = 8$~TeV. No significant deviation from the Standard Model background expectation is observed, and lifetime-dependent upper limits on $R$-hadrons and chargino production are set. Gluino $R$-hadrons with 10~ns lifetime and masses up to 1185~GeV are excluded at 95$\%$ confidence level, and so are charginos with 15~ns lifetime and masses up to 482~GeV.
}

\PreprintJournalName{European Physics Journal C}

\begin{document}
\title{Search for metastable heavy charged particles
with large ionisation energy loss in \emph{pp} collisions at \boldmath{$\sqrt{s} = 8$}~TeV  using the \mbox{ATLAS} experiment}
\titlerunning{Search for metastable heavy charged particles with large ionisation energy loss}
\author{The ATLAS Collaboration}
\authorrunning{The ATLAS Collaboration}
\institute{CERN, 1211, Geneva 23, Switzerland}
\date{Received: date / Revised version:}
%
%
\abstract{
Many extensions of the Standard Model predict the existence of charged heavy long-lived particles, such as $R$-hadrons or charginos.
These particles, if  produced at the Large Hadron Collider, should be moving non-relativistically and are therefore identifiable through the measurement of an anomalously large specific energy loss in the \mbox{ATLAS} pixel detector. Measuring heavy long-lived particles through their track parameters in the vicinity of the interaction vertex provides sensitivity to metastable particles with lifetimes from 0.6~ns to 30~ns.
A search for such particles with the \mbox{ATLAS} detector at the Large Hadron Collider is presented, based on a data sample corresponding to an integrated luminosity of $\lumi$~fb$^{-1}$ of  $pp$ collisions at $\sqrt{s} = 8$~TeV. No significant deviation from the Standard Model background expectation is observed, and lifetime-dependent upper limits on $R$-hadrons and chargino production are set. Gluino $R$-hadrons with 10~ns lifetime and masses up to 1185~GeV are excluded at 95$\%$ confidence level, and so are charginos with 15~ns lifetime and masses up to 482~GeV.
}
\maketitle


\section{Introduction} \label{sec:Introduction}
  The main motivation for heavy long-lived particle (LLP) searches at the Large Hadron Collider (LHC) arises from proposed solutions to the gauge hierarchy problem~\cite{Susskind:1982mw}, which typically involve previously unseen particles at the TeV mass scale. Hadronising LLPs are anticipated in a wide range of physics models that extend the Standard Model (SM). For example, these particles appear in both $R$-parity-conserving~\cite{Fairbairn:2006gg,Kolda:1997wt,Baer:1998pg,Gates:1999ei,Mafi:1999dg,Chertok:2001ja,Giudice:2004tc,ArkaniHamed:2004yi} and $R$-parity-violating\\~\cite{Dreiner:1997uz,Berger:2003kc,Barbier:2004ez} supersymmetry (SUSY) and in universal extra dimensions theories~\cite{Feng:2003nr,Shah:2006gs}.
  
These particles can be stable, or metastable;\footnote{In this article, particles that do not decay in the active detector volume are considered stable, and those that decay inside this volume are considered metastable.} the lifetime of these metastable particles may depend on the mass splitting with the lightest SUSY particle or on the size of any $R$-parity-violating coupling~\cite{Randall:2011}. 

LLPs produced at the LHC are expected to be slow ($\beta$ significantly below 1) and, therefore, 
should have specific ionisation higher
than any SM particle at high momenta. The \mbox{ATLAS} detector~\cite{:2008zzm} has a number of subsystems 
able to measure the velocity of charged particles.
The pixel detector~\cite{Aad:2008zz} provides measurements of ionisation energy loss ($\dedx$) whereas the calorimeters 
and the muon spectrometer give a direct measurement of the time of flight for particles traversing them.
A search for stable LLPs has been performed with the ATLAS detector~\cite{ATLAS:2013Rhad} with 4.7~fb$^{-1}$ of $\sqrt{s} = 7$~TeV proton--proton ($pp$) collisions using both the full detector information as well as the pixel detector information alone, and has been recently updated with the entire $\sqrt{s} = 8$~TeV dataset~\cite{Shlomit_2015}, but without a pixel-only analysis.
The CMS Collaboration has recently published~\cite{CMS:2013LLP} an analysis searching for stable LLPs based on the measurement of $\dedx$, $\beta$, and on muon identification. 
In CMS, a search for metastable LLPs has been carried out by looking for secondary vertices~\cite{CMS_sec_jets}, or disappearing tracks~\cite{CMS_disappear}.
A displaced vertex search performed by the ATLAS Collaboration~\cite{Aad:2015rba} sets limits on metastable particles in a number of scenarios.
Limits on chargino production, from an analysis searching for disappearing tracks, have also been published by the ATLAS Collaboration~\cite{Shimpei}.

The analysis described in this article has sensitivity to metastable particles if they have unit charge and their track length before decay is more than 45~cm in the radial direction, so that they can be measured in the first few layers of the ATLAS tracker.
This measurement does not depend on the way the LLP interacts in the dense calorimeter material nor, to a first approximation, on the LLP decay mode. 
It can therefore address many different models of New Physics, especially those predicting the production of metastable heavy particles with \emph{O}(ns) lifetime at LHC energies, such as mini-split SUSY~\cite{Arvanitaki:2013,SpreadSUSY:2012} or anomaly-mediated supersymmetry breaking (AMSB) models~\cite{ASMB1:Giudice1998,ASMB2:Randall1999}.
A metastable gluino with a mass of 1~TeV would be compatible with the measured Higgs boson mass according to mini-split SUSY models, which also predict squark masses of $10^3$--$10^5$~TeV, therefore making the gluino the only observable strongly produced SUSY particle at LHC energies. 
In AMSB models, SUSY breaking is caused by loop effects and the lightest chargino can be only slightly heavier than the lightest neutralino, resulting in a heavy charged particle that can be measured before
decaying into very low energy SM particles and a neutralino.

Results are presented in the context of SUSY models assuming the existence of $R$-hadrons~\cite{Farrar:1978xj} formed from a long-lived coloured sparticle (squark or gluino) and light SM  quarks or gluons, and in AMSB models for the case of long-lived charginos.

The paper  is organised as follows.
After a brief description of the experiment (Section~\ref{sec:Experiment}) and of the measurement strategy (Section~\ref{sec:Strategy}), the simulation of the signal processes is described (Section~\ref{sec:Simulation}). The triggering strategy is then summarised and the trigger efficiency is calculated for  $R$-hadrons and charginos (Section~\ref{sec:Trigger}), after which the event selection is defined and motivated (Section~\ref{sec:Selection}).
The data-driven background estimation is then described (Section~\ref{sec:Background}) and the systematic uncertainties are presented and discussed (Section~\ref{sec:Systematics}).
Finally, after a brief description of the statistical method used to extract lifetime-dependent limits on $R$-hadron and chargino production cross sections and masses, the results are reported in Section~\ref{sec:results}.


%
\section{ATLAS detector and pixel \boldmath{$\dedx$} measurement}\label{sec:Experiment}
The ATLAS detector\footnote{ATLAS uses a right-handed coordinate system with its origin at the nominal interaction point in the centre of the detector and the $z$-axis coinciding with the axis of the beam pipe. The $x$-axis points from the interaction point to the centre of the LHC ring, and the $y$-axis points upward. Cylindrical coordinates ($r$, $\phi$) are used in the transverse plane, $\phi$ being the azimuthal angle around the beam pipe. The pseudorapidity is defined in terms of the polar angle $\theta$ as $\eta = -\ln \tan(\theta/2)$.} consists of a tracker surrounded by a solenoid magnet for measuring the trajectories of charged particles, followed by calorimeters for measuring the energy of particles that have electromagnetic or strong interactions with matter, and a muon spectrometer. 
The muon spectrometer is immersed in a toroidal magnetic field and provides tracking for muons, which have typically passed through the calorimeters. The detector is hermetic within its $\eta$ acceptance and can therefore measure the missing transverse momentum (whose magnitude is denoted by $\MET$) associated with each event. A complete description of the ATLAS detector can be found elsewhere~\cite{:2008zzm}.
The tracker is made of three detector systems.  Starting from the solenoid magnet and moving toward the beam collision region one finds a $\approx$400-thousand-channel transition radiation tracker~\cite{ATLAS_TRT}  followed by a  $\approx$6-million-channel silicon microstrip detector~\cite{ATLAS_SCT}, and finally a  $\approx$80-million-channel pixel detector.  
The pixel detector is crucial for this measurement and is described in more detail below.

As the innermost sub-detector in ATLAS, the silicon pixel detector provides at least three precision measurements for each track in the region $|\eta| <2.5$ at radial distances of 5 to 13~cm from the LHC beam line. 
At normal incidence, the average charge released by a minimum-ionising particle (MIP) in a pixel sensor is $\approx$~20000~\textit{e}$^{-}$ and the charge threshold is set to 3500$\pm$40~\textit{e}$^{-}$ for each pixel. Signals above this threshold are time-stamped within one beam crossing; the hit efficiency under these conditions exceeds 99$\%$.
When detector data are read out, the time over threshold (ToT), i.e. 
the length of time for which the signal is above the threshold, is digitised with 8 bits.
The ToT is proportional to the ionisation charge~\cite{BaBar_ToT} and its maximum value corresponds to 8.5 times the average charge released by a MIP track normal to the silicon detectors and leaving all of its ionisation charge on a single pixel. If this value is exceeded, the signal is lost. 

The charge released by a track crossing a layer of the pixel detector is rarely contained within just one pixel. Neighbouring pixels are thus joined together to form clusters and the charge of a cluster is calculated by summing up the charges of all pixels after calibration corrections. The specific energy loss ($\dedx$) is defined as an average of the individual cluster ionisation measurements (charge collected in the cluster, corrected for the track length in the sensor), for the clusters associated with the track. To reduce the Landau tails, the average is evaluated after having removed the highest $\dedx$ cluster and amounts to~$1.24 \pm 0.19$ MeV/g cm$^{2}$ for a MIP~\cite{dedxnote}. The minimum measurable $\beta\gamma$ with the $\dEdx$ method is $\approx$0.3 for particles with unit charge and is determined by the ToT overflow in any of the pixels in a cluster.


\section{Measurement strategy}\label{sec:Strategy}
Charged massive LLPs are expected to interact with matter  following the Bethe--Bloch distribution according to their $\beta\gamma$.
The mass of the LLPs can be obtained by fitting their specific energy loss and momentum to an empirical Bethe--Bloch distribution in the range $0.3< \beta\gamma<1.5$.
This range overlaps with the expected average $\beta\gamma$ of LLPs produced at the LHC, which decreases from 2.0 to 0.5 for an increase in particle mass from 100~GeV to 1600~GeV.
The parametric function describing the relationship between the most probable value of the energy loss ($ \dEdx_{\mathrm{MPV}}$) and $\beta\gamma$ is~\cite{dedxnote}:
\begin{equation}
 \dEdx_{\mathrm{MPV}}(\beta\gamma) = {p_1\over \beta^{\,p_3}}\ln[1+(|p_2|\beta\gamma)^{p_5}]-p_4.
 \label{bbfun}
\end{equation}
The $p_i$ calibration constants 
have been measured~\cite{dedxnote} using low-momentum pions, kaons
and protons reconstructed in ATLAS and their values are monitored by checking the stability of the proton mass measurement as a function of time.
The calibration is found to be stable at the 1\% level for all data-taking conditions and detector settings. 

\begin{figure}
\resizebox{0.50\textwidth}{!}{%
\includegraphics{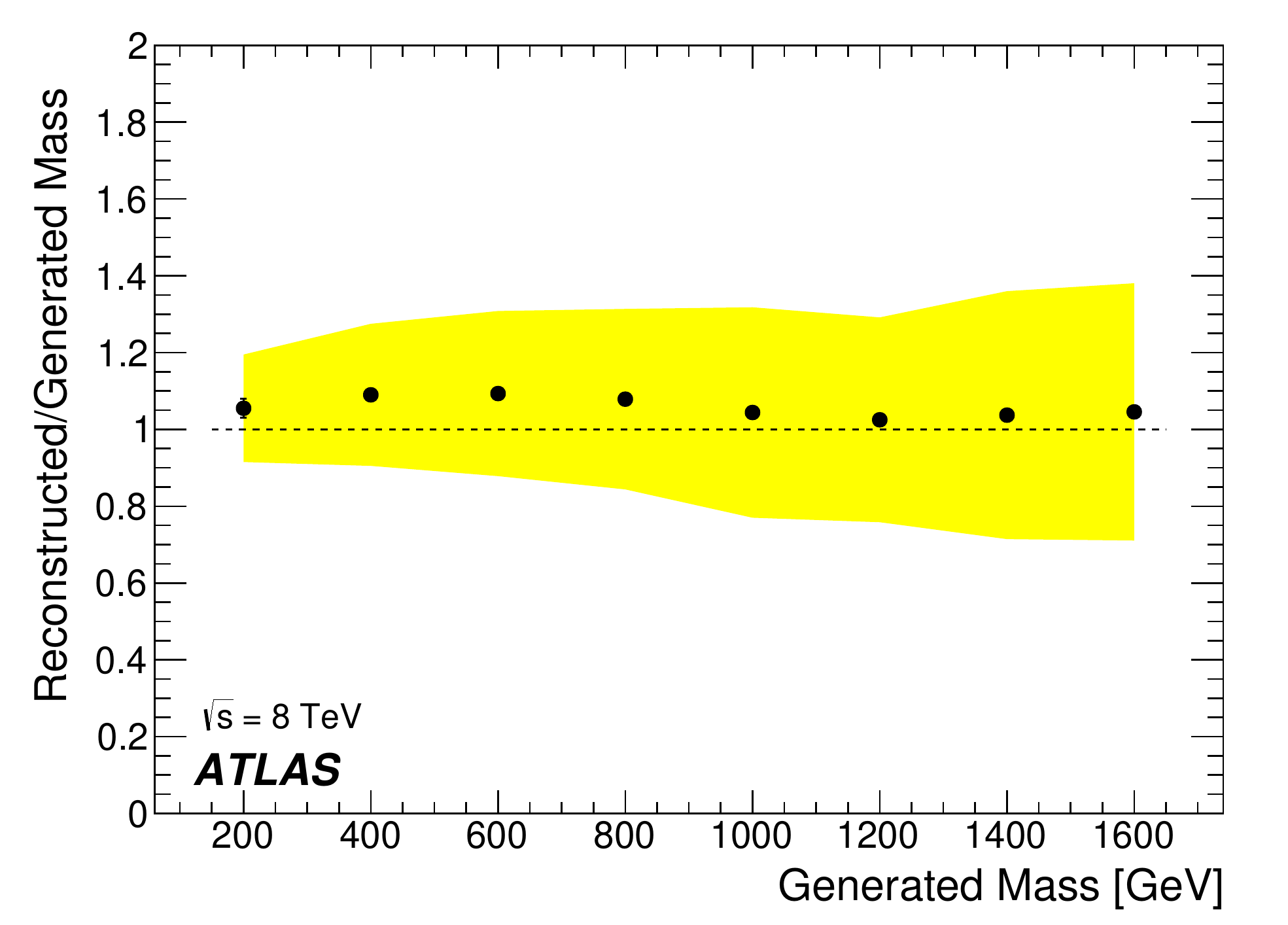}
}
\caption{Ratio of the reconstructed mass, computed as the most probable value of a fit to a Landau distribution convolved with a Gaussian, to the generated mass, as a function of the generated mass for stable gluino $R$-hadrons. The yellow band is the half-width at half maximum of the reconstructed mass distribution normalised to the generated mass.}
\label{fig:RecvsGen}    
\end{figure}

Given a measured value of  $\dEdx$, the mass estimate $m$  is obtained from Eq.~(\ref{bbfun}) by numerically solving
the equation $\dEdx_{\mathrm{MPV}}(p/m) = \dedx$ for the unknown $m$.
Figure~\ref{fig:RecvsGen} shows the ratio of the reconstructed mass to the generated mass for simulated $R$-hadrons with masses up to 1600~GeV. 
The overestimation of the reconstructed mass for heavy particles is due to the pion scattering model assumed in the track reconstruction and momentum measurement. A $\approx$3\% rescaling of the measured mass is then applied in the analysis.
The half-width at half maximum of the reconstructed mass distribution increases with the mass value.
This is due to the momentum measurement uncertainty dominating the mass resolution above masses of 200~GeV.

The measurement strategy consists of looking for an excess of events, compatible with the expected measurement resolution, in the mass distribution for particles that are selected as LLP candidates. 
These should appear as high transverse momentum ($\pt$) isolated particles with large $\dEdx$.
Since there is no trigger based on these observables, events of interest were selected using the lowest-threshold unprescaled calorimetric $\MET$ trigger. 
For signal events, the $\MET$ originates from jets from QCD initial-state radiation (ISR) and, whenever relevant, by the LLP decays to undetected neutralinos. The trigger efficiency for all the LLPs searched for with this analysis is shown in Section~\ref{sec:Trigger}.


\section{Simulation of signal}\label{sec:Simulation}
A number of Monte Carlo (MC) simulated signal samples are used in this analysis to determine the expected efficiencies and to estimate the systematic uncertainties. 
A description of the simulation techniques is presented below for both stable and metastable $R$-hadrons and charginos.
All simulated events are processed through the  \textsc{Geant4} ~\cite{Agostinelli:2002hh} standard ATLAS simulation~\cite{:2010wqa} and digitisation, followed by event reconstruction. This includes a realistic description of additional $pp$ interactions in the same or neighbouring bunch crossings (pile-up). In order to take into account residual discrepancies with data, the simulated pile-up distribution is scaled to that observed during the 2012 data-taking period.

\subsection{Stable \boldmath{R}-hadrons}\label{sec:R-hadrons}

Pair production of gluinos with masses between 100 and 1700~GeV is simulated in \textsc{Pythia 6.4.27}~\cite{Sjostrand:2006za} with the AUET2B~\cite{ATLAS_tunes_2011-014} set of MC tuneable parameters and the CTEQ6L1~\cite{ATLAS_PDF_set} parton distribution function (PDF) set, incorporating dedicated hadronisation routines~\cite{Kraan:2004tz} to produce final states containing $R$-hadrons. 
Additional samples of gluinos with some representative mass values are generated using {\sc MadGraph5}~\cite{Madgraph} 1.3.33. 
Since the {\sc MadGraph} samples are generated with an additional outgoing parton in the matrix element they provide a more accurate description of ISR and thus a more accurate distribution of the transverse momentum of the gluino--gluino system. The gluino samples simulated with \textsc{Pythia6} are reweighted to match this gluino--gluino system $\pt$ distribution obtained from the {\sc MadGraph} samples. 
The cross sections are calculated to next-to-leading order in the strong coupling constant (NLO), including the resummation of soft-gluon emission at next-to-leading-logarithmic accuracy (NLO+ NLL)~\cite{Beenakker:1996ch,Kulesza:2008jb,Kulesza:2009kq,Beenakker:2009ha,Beenakker:2011fu}. 
The nominal cross section is calculated assuming a squark mass of 10~TeV. The uncertainty is taken from an envelope of cross-section predictions using different PDF sets and factorisation and renormalisation scales, as described in Ref.~\cite{Kramer:2012bx}.

Simulated $R$-hadron events are passed through a full detector simulation, where interactions with matter are handled by dedicated \textsc{Geant4} routines based on different scattering models.
The model described in Refs.~\cite{Kraan:2004tz,Mackeprang:2006gx}, and hereafter referred to as the generic model, imposes few constraints on the allowed stable states. This is the only model where doubly charged $R$-hadrons are predicted, with a production probability of 0.1\%. 
 Hadronic scattering is described through a purely phase-space-driven approach.
A second model, referred to as the Regge model in the following, employs a triple-Regge formalism to describe hadronic scattering and describes $R$-hadrons containing gluinos according to Ref.~\cite{deBoer:2007ii}. 
More recent models for the hadronic scattering of gluino $R$-hadrons predict that the majority would be electrically neutral after just a few hadronic interactions. 
The third scenario considered here belongs to this family and is based on bag-model calculations presented in Ref.~\cite{Farrar:2010ps}. This is referred to as the intermediate model.
The probability for a gluino to form a gluon--gluino bound state, based on a colour-octet model, is assumed to be 10$\%$~\cite{Fairbairn:2006gg}.
Results are presented for the generic (Regge) model for gluino (squark) $R$-hadrons. Variations resulting from the use of a model different from the nominal one are taken into account as systematic uncertainties on the signal efficiency.

\subsection{Metastable \boldmath{R}-hadrons}\label{sec:metaRH}

The simulation of metastable gluino-based $R$-hadron samples is performed in a similar way to that of the stable $R$-hadrons, as described in Section~\ref{sec:R-hadrons}. The gluinos within $R$-hadrons are required to decay via the radiative process
$\tilde{g}\to g\tilde{\chi}_1^0$ or  $\tilde{g}\to q\bar{q}\tilde{\chi}_1^0$, using \textsc{Pythia6}. Decays involving $t\bar{t}$  pairs are treated separately. 
Gluino masses between 400~GeV and 1400~GeV are simulated, with the neutralino mass either fixed to 100~GeV or set to $m(\tilde{g}) - 100$~GeV (or $m(\tilde{g}) - 480$~GeV for $t\bar{t}$ channels). The gluino lifetime is varied from 0.1~ns to 10~ns. 

Results of this search are presented as a function of the $R$-hadron lifetime. 
Signal MC samples with different lifetimes are obtained, starting from those simulated with a fixed value of the lifetime, by applying event weights, such that the distribution of the proper lifetime in the modified sample corresponds to the chosen new value of the mean lifetime. 
The event weight $w$  is given by:
\begin{equation}
w(\tau_{RH}) = \prod_i^{n_{RH}}\frac{\tau_0}{\tau_{RH}}\exp\Bigg[-t_i\Bigg(\frac{1}{\tau_{RH}}-\frac{1}{\tau_0}\Bigg)\Bigg],
\end{equation}
where $n_{RH}$, $\tau_0$ and $t_i$ are the number of $R$-hadrons in the event, the $R$-hadron mean lifetime as set by the simulation, and the proper lifetime of the $i^{th}$ $R$-hadron, respectively. The modified mean lifetime obtained after the reweighting procedure is indicated by $\tau_{RH}$.

\subsection{Stable charginos}\label{sec:charginos}

A chargino may be stable in a simplified scenario where the mass difference between the chargino and neutralino is less than 140 MeV so that the chargino decay to a pion and a neutralino is kinematically suppressed.
Samples with long-lived charginos are generated using \textsc{Herwig++ 2.6.3}~\cite{Bahr:2008pv} along with the UEEE3~\cite{Gieseke:2012} tune and the CTEQ6L1 PDF set. The chargino mass is varied between 100~GeV and 800~GeV.
 The chargino is forced to remain stable and the other particles in the model are set to be too heavy to be produced in $\sqrt{s}=8\tev$ $pp$ collisions. Signal cross sections are calculated to NLO using PROSPINO2~\cite{Beenakker:1999xh}. They are in agreement with the NLO+NLL calculations within $\sim$2\%~\cite{Fuks:2012qx,Fuks:2013vua,Fuks:2013lya}.
The total cross section is dominated by direct production of $\tilde{\chi}_1^0\tilde{\chi}_1^\pm$ pairs ($\sim67\%$), and by $\tilde{\chi}_1^+\tilde{\chi}_1^-$ pairs ($\sim30\%$). The relative proportion of these two production mechanisms was checked and found to be constant at the 1\% level over the considered chargino mass range.

\subsection{Metastable charginos}\label{sec:charginos_meta}

Samples with metastable charginos are produced similarly to the samples of stable charginos. The mean lifetime of the chargino is set to a given value ($\tau_{\tilde{\chi}_1^\pm}$), and charginos are forced to decay into $\tilde{\chi}^0_1 + \pi^\pm$ in the \textsc{Geant4} simulation following an exponential decay with lifetime $\tau_{\tilde{\chi}_1^\pm}$ in the chargino rest frame. 
The chargino--neutralino mass splitting is set to 140~$\mev$. 
Samples with $\tau_{\tilde{\chi}_1^\pm}$ = 1, 5, 15 and 30~ns are generated for different mass points.  
To reduce the use of simulation resources, all the samples were generated with a jet filter requiring at least one generator-level jet with $\pt >70$~GeV and $|\eta|<5$. This choice was optimised for the first metastable chargino search~\cite{Shimpei}.
The present analysis does not make any explicit requirement on the energy of the jets, while it requires missing transverse momentum in the event. There is a strong correlation between these two requirements. The residual bias due to the jet filter is evaluated and assigned as a systematic uncertainty.


\section{Candidate selection}\label{sec:Selection}

\subsection{Trigger}\label{sec:Trigger}

\begin{figure}
\resizebox{0.50\textwidth}{!}{%
    \includegraphics{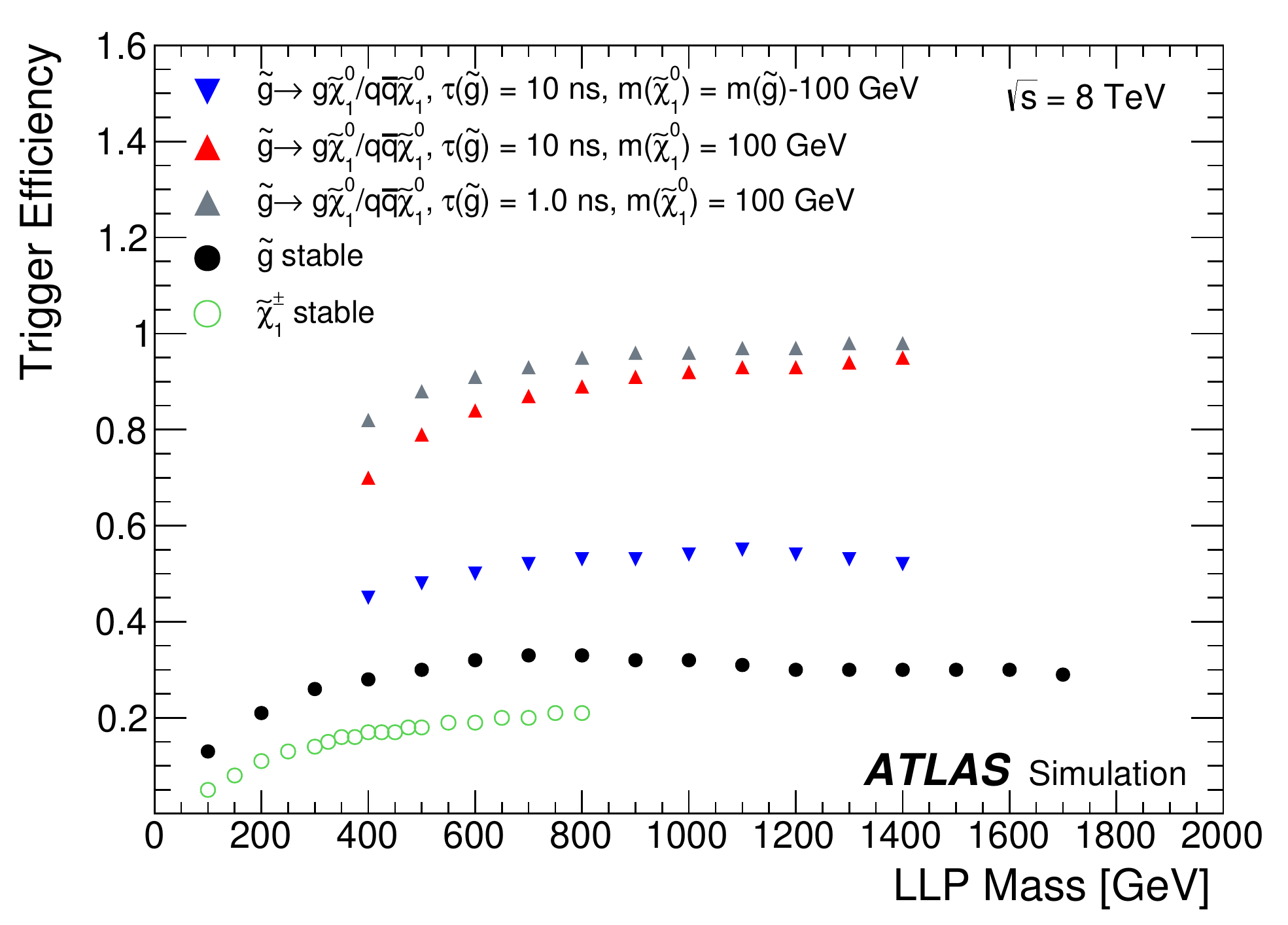} 
}
  \caption{Efficiency for the calorimetric $\MET>80$~\rm{GeV} trigger as a function of the $R$-hadron mass or of the chargino mass. Separate curves are also shown for the stable and metastable cases. Only statistical uncertainties are shown, which are too small to be visible on these graphs.}
\label{fig:TriggerEfficiency_vs_mass}
\end{figure}

Events are selected by  the $\MET>80$~GeV trigger, the lowest-threshold $\MET$ trigger that remained unprescaled throughout the 2012 data taking.
This is based uniquely on the energy deposited in the calorimeters. 
Figure~\ref{fig:TriggerEfficiency_vs_mass} shows the efficiency of this trigger as a function of the $R$-hadron or chargino mass.

The decay of each LLP to jets and a neutralino occurring within the ATLAS active volume contributes to the transverse momentum imbalance and leads to higher trigger efficiency in the metastable cases.
In the case of metastable $R$-hadrons, the $\MET$ also depends  on the mass of the neutralino produced in the decay itself. 
The two $R$-hadrons tend to be emitted back-to-back and the same holds for the two heavy neutralinos, which therefore approximately balance  $\MET$. If neutralinos are light, high-$\pt$ jets coming from $R$-hadron decay also contribute to the transverse momentum inbalance.
For this reason, the calorimetric $\MET$ is significantly larger when the neutralino is light. 
The $\MET$  also depends on the lifetime of the parent particle as this defines the fraction that decay  before the calorimeter and therefore affects $\MET$. 
If the lifetime is shorter than 1~ns, the decay happens very close to the primary vertex and the calorimetric $\MET$ does not depend very much on the lifetime. On the other hand, if the lifetime is long enough, the decay may happen beyond the calorimeter region and, therefore, the calorimetric $\MET$ is close to the stable case. The 10~ns lifetime is an intermediate case, as decays happen mainly, but not exclusively, in the tracker region. 


\subsection{Offline selection}

This search is based on a sample of well-measured high-$\pt$ isolated tracks in events with large missing transverse momentum. 
The data sample considered in this analysis was collected with tracking detectors, calorimeters, muon chambers and magnets fully operational and corresponds to a total integrated  luminosity of $\lumi$~fb$^{-1}$ with an uncertainty of $\pm$2.8\% measured using beam separation scans following the technique described in Ref.~\cite{ATLAS:lumi}. 

The first step in the selection is the confirmation that the event has sufficient $\MET$. The $\MET$ variable computed using the offline reconstruction~\cite{ATLAS:etmiss,ATLAS:etmiss1}, which uses refined calorimetric information and includes the contributions of the energy of the muons, must exceed 100~GeV. Candidate events are then required to have  at least one primary vertex with a minimum of five tracks with $\pt > 0.4~\GeV$. There must be at least one track associated with this primary vertex,\footnote{A track is associated with the primary vertex if its longitudinal and transverse impact parameters are less than 1.5~mm. The primary vertex is chosen as that with the highest sum of $\pt^2$ associated with it.} with at least three pixel hits, measured over 45~cm in the radial direction, and with transverse momentum $\pt > 80~\GeV$ and $|\eta| \leq 2.5$. The set of requirements described above, including the trigger requirement, defines the preselection entry in Tables~\ref{tab:cutflowData}~and~\ref{tab:cutflowGluino1000}. 

The following additional requirements must be satisfied by at least one of the preselected candidate tracks in order to select the event.

The track must be isolated. A track is considered isolated if  its distance $\Delta R = \sqrt{ (\Delta\eta)^2 + (\Delta\phi)^2}$ 
to any other track associated with the primary vertex and with $\pT \geq 1$~GeV is greater than $0.25$. About 70\% of these isolated tracks are  high-$\pt$ leptons originating from $W$ boson production.

The track must not be identified as an electron~\cite{ATLAS:electrons} as LLPs can very rarely ($<$1\%)  be identified as electrons. The selected tracks are required not to match any reconstructed electron within $\Delta R \le 0.01$. 

The track must have momentum $p >150~\GeV$ and the relative uncertainty on the momentum $\sigma_{p}/p < 50\%$. The first requirement improves the signal-to-background ratio, while the second ensures good mass resolution.

The track must not be a muon originating from a $W$ boson decay. Muons
 cannot be simply identified and rejected at this stage, as hypothetical 
very long-lived particles would often  be mis-identified as muons in 
the detector. Therefore, to reject muons from a $W$ boson decay, a requirement on transverse mass\footnote{$m_{\rm T} = \sqrt{2 \pt \MET (1-\cos(\Delta\phi (\MET,\rm{track}) )}$ } ($m_{\rm T}>$130~GeV) is applied. 
According to simulation, this requirement reduces the  fraction of $W$ boson events in the data sample to $\sim 40$\%.  

The selected track is required to have specific ionisation measured by the pixel detector larger than 
$1.800 - 0.034 |\eta| + 0.101 \eta^2 - 0.029 |\eta|^3 $~MeV/g\,cm$^{2}$.
This requirement corrects the slight $|\eta|$ dependence~\cite{Bichsel88} of the $\dEdx$ variable and selects  $\sim 1.3$\% of the tracks in the data independently of the pseudorapidity region. The selection cut chosen is the lowest with mass-discriminating power (below this, the $\dedx$ values of all particles are too close to the MIP value, irrespective of their masses). 

The above requirements complete the selection for the stable particle search. One additional requirement is applied to improve the sensitivity for the metastable case.
The highly ionising particles can be matched with reconstructed jets~\cite{ATLAS:jets} or muons~\cite{ATLAS:muons}. 
Out of 85 candidates, 57 are geometrically matched to muons ($\Delta R \le 0.01$) and 26 are $\Delta R \le 0.07$ from a jet. The other two candidates have no signals in the calorimeters or muon system in the vicinity of the LLP. If the LLPs are stable, they are usually reconstructed as muons. If the heavy particles are not stable, the matching with muons becomes much more rare, in particular for particles with a lifetime of~ \emph{O}(ns).
In the search for metastable particles a muon veto is applied, and tracks that are matched with a muon are rejected.

Finally, if more than one track per event passes all requirements, the highest-$\pt$ candidate is chosen, in order not to bias the distribution of the variables and to allow for proper normalisation in the background estimate. 

Table~\ref{tab:cutflowData} shows the number of events in data and for an example gluino $R$-hadron signal for the different selection criteria. In data 85 events are selected before and 28 after the muon veto. None of the events has more than one selected track per event.

\begin{table}
\begin{center}
    \caption[]{Observed data event yields at different steps of the selection procedure compared with the expected number of events for 1000~GeV  $R$-hadrons decaying, with a 10~ns lifetime, to $g/q\bar{q}$ plus a light neutralino of mass $m(\tilde{\chi}^0_1)=100$~GeV. The simulated yields are normalised to a total integrated luminosity of $\lumi$~fb$^{-1}$ and their statistical uncertainty is also shown. See text for details.

    \label{tab:cutflowData}}
  \vspace{0.3cm}
 \begin{tabular}{| l | c c|}
    \hline
   {\bf } & {\bf Selected events  } &{\bf Expected events}\\
   {\bf Requirement} & {\bf  (data) } &{\bf ($\tau = 10$~ns)}\\

    \hline
Preselection      &      543692 &  112$\pm$3\\                
Isolation		&	88431 &102$\pm$3\\	
Electron veto	&	60450 & 102$\pm$3\\	
High-$p$ 		&	35684 & 91$\pm$3\\	
High-$m_{\rm T}$	&	6589	  & 75$\pm$2\\	
Ionisation		&	85	&  68$\pm$2\\	
    \hline
Muon veto	&	28	&  62$\pm$2\\
    \hline
    \end{tabular}
    \end{center}
\end{table}

Table~\ref{tab:cutflowGluino1000} shows the yields for the same event selection as in Table~\ref{tab:cutflowData}, but applied to simulated signal events with 1000~GeV gluino $R$-hadrons that are either stable, or otherwise decay to $g/q\bar{q}$ plus a light neutralino of mass $m(\tilde{\chi}^0_1)=100$~GeV, and with a 1~ns lifetime.

\begin{table}
\begin{center}
    \caption[]{Expected number of events at different steps of the selection procedure for 1000~GeV gluino $R$-hadrons decaying, with a 1~ns lifetime, to $g/q\bar{q}$ plus a light neutralino of mass $m(\tilde{\chi}^0_1)=100$~GeV, and for stable $R$-hadrons. The simulated yields are normalised to a total integrated luminosity of $\lumi$~fb$^{-1}$ and their statistical uncertainty is also shown. See text for details.
    \label{tab:cutflowGluino1000}}
  \vspace{0.3cm}
  \begin{tabular}{| l | c c|}
    \hline
   {\bf } & {\bf Expected events  } &{\bf Expected events}\\
   {\bf Requirement} & {\bf  ($\tau = 1$~ns) } &{\bf (stable)}\\

    \hline
Preselection      &     19.5$\pm$1.1           &      62$\pm$2       \\
Isolation 		& 	9.3$\pm$0.7 	     & 	      55$\pm$2 	\\
Electron veto 	& 	9.3$\pm$0.7 	    &         54$\pm$2 \\
High-$p$  	        & 	4.4$\pm$0.5 		&   52$\pm$2	\\
High-$m_{\rm T}$ 	& 	3.9$\pm$0.4 		& 	43$\pm$1 \\
Ionisation 		& 	3.0$\pm$0.4		& 	37$\pm$1 \\
    \hline
Muon veto 	& 	2.9$\pm$0.4		& 	 $-$	\\
    \hline
    \end{tabular}
    \end{center}
\end{table}

\begin{figure}
\resizebox{0.50\textwidth}{!}{%
    \includegraphics{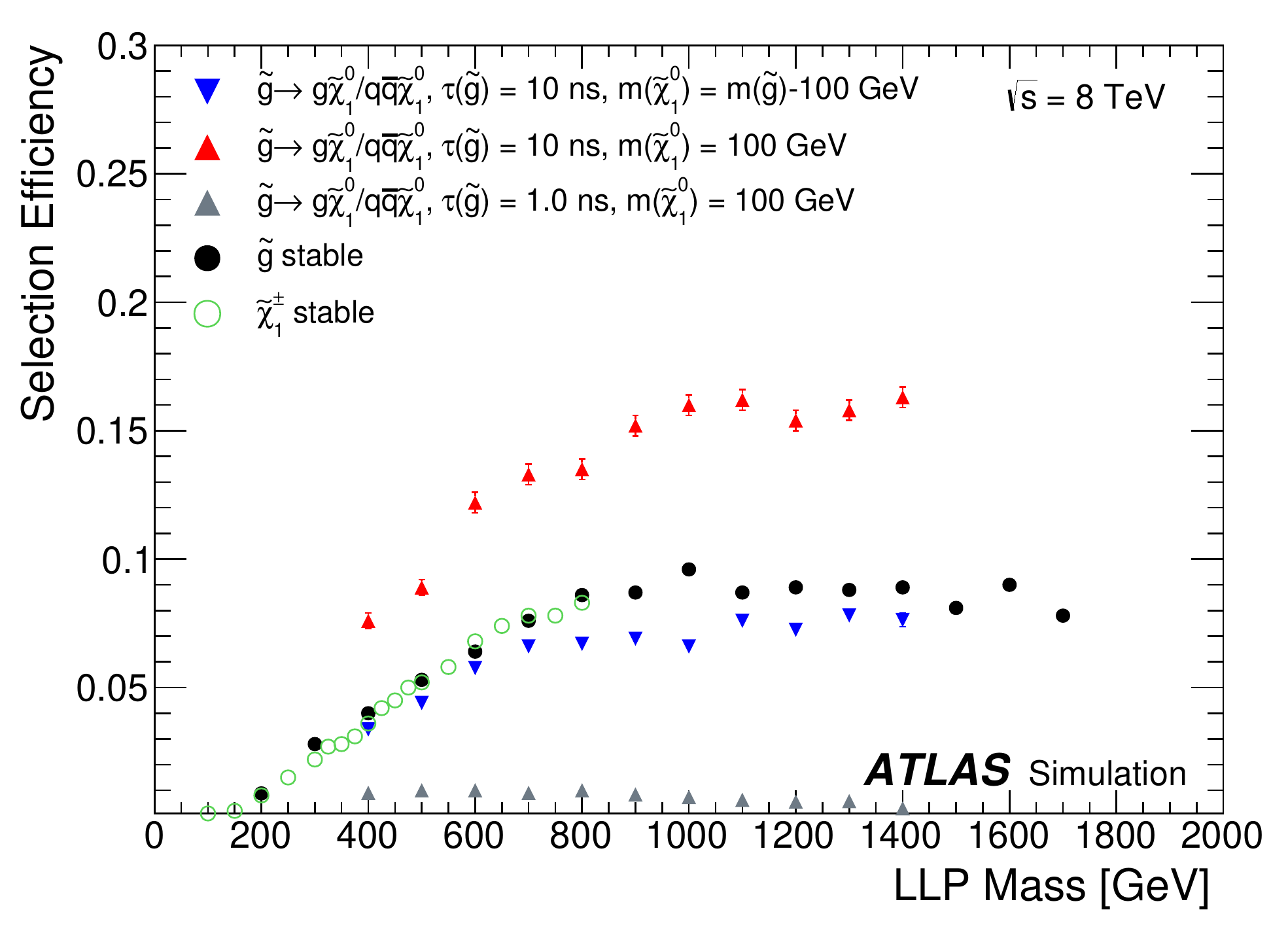}
}
\caption{Total selection efficiencies for some MC samples as a function of the $R$-hadron mass or of the chargino mass. Both stable and some metastable cases are shown. Only statistical uncertainties are included.}
\label{fig:TotalEfficiency}
\end{figure}

Figure~\ref{fig:TotalEfficiency} shows the overall signal efficiencies for a representative set of simulated signal samples to which the full selection procedure is applied.

When the LLPs decay inside the ATLAS active volume, $\MET$ increases and trigger and offline $\MET$ selection becomes more efficient than for the stable case. However, as  the lifetime decreases, the probability to reconstruct a track segment in the silicon detectors decreases dramatically. At a mass of 1000~GeV, these two effects give a total efficiency of $\approx$15\% for the 10~ns lifetime samples and $\approx$1\% for the 1~ns samples, while for stable particles the efficiency has intermediate values of $\approx$7\%.


\section{Background estimation}\label{sec:Background}

In order to estimate the background, a data-driven approach is used. The method uses data to fit the distributions of key variables, taking into account their inter-dependence, and then to generate a large  random sample of background events based on the same distributions.
The choice of the control samples takes into account the measured correlations between the variables used: $p$, $\dEdx$ and $\eta$.
The $\dEdx$ dependence on the path length in the sensor is not linear~\cite{Bichsel88}, but depends on $\eta$, increasing by $\sim$10\% from central to high $|\eta|$. The $\dEdx$ also depends on the particle $\beta\gamma$ via the Bethe--Bloch formula and therefore on its momentum, until the Fermi plateau is reached. Finally $p$ and $\eta$ depend on event kinematics, as high-momentum tracks are more likely to be produced at high $|\eta|$ values.

Two samples are constructed to describe the distributions of the key variables.
Both selections use the full data sample, but with requirements minimising the possible contamination by signal events.
These control region samples are the same for the searches for stable and metastable particles, except for the rejection of track candidates geometrically matched with spectrometer muons in the latter case.
   
   A first sample (\textit{CR1}) is selected by applying all the selections described in Section~\ref{sec:Selection}
   except for the high ionisation requirement, which is instead inverted to 
   ensure  orthogonality with the search sample. Otherwise, the kinematic properties and overall event characteristics are expected to be similar to the signal region.
 With this selection, $\sim$6000 background events are kept in  \textit{CR1} with signal contamination of less than 0.6\%.

\begin{figure}
\resizebox{0.50\textwidth}{!}{%
	   \includegraphics{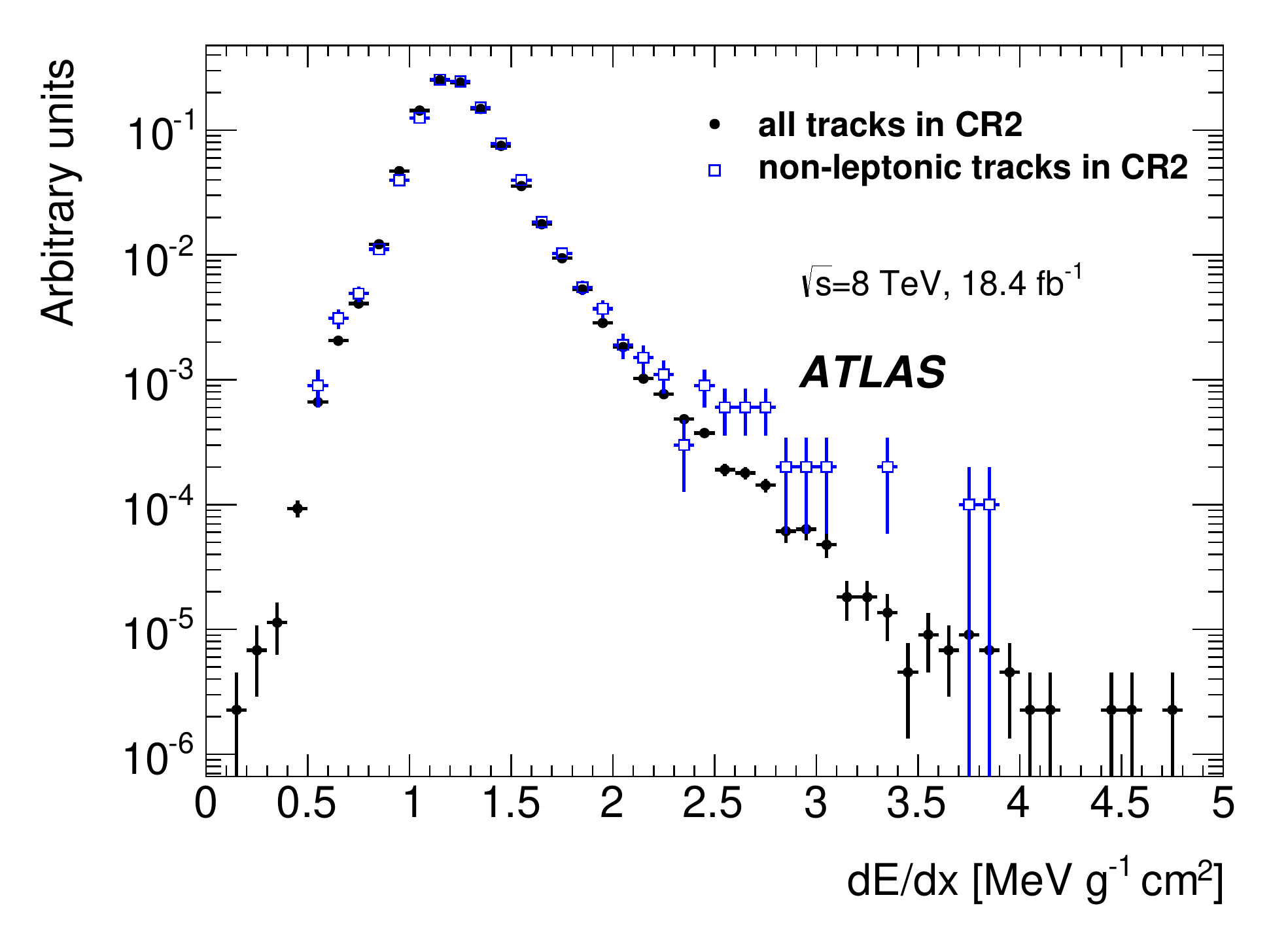}
 }
\caption{Ionisation distribution of all the  \textit{CR2} tracks (filled circles), and those not matched to a reconstructed muon (open squares). The two distributions are normalised to their total number of entries.}
\label{fig:parametrized_dEdx_alleta}
\end{figure}

A second sample (\textit{CR2}) is used to obtain the $\dEdx$ templates and is selected by inverting the $\MET$ requirement ($\MET<$100~GeV), while keeping all the other selection requirements unchanged. This procedure ensures a negligible signal content in the sample (signal contamination of less than 0.02\%), and that the selected tracks are in the same kinematic ranges of momentum and pseudorapidity as the search sample.
With this selection, $\sim440000$  background events are kept in  \textit{CR2}, the majority ($\approx$ 96\%) being matched with muons.
The ionisation of all the \textit{CR2} tracks and of the subset with the muon veto applied is shown in Figure~\ref{fig:parametrized_dEdx_alleta}. 
In their bulk the distributions are very similar, thus the larger \textit{CR2} sample is also used for the metastable search with the muon veto. Nevertheless a larger tail, likely due to $e^+e^-$ pair production, is seen at high $\dEdx$ for the tracks with muon veto, and this difference is accounted for in the systematic uncertainty evaluation.

A large background sample consisting of five million \{$p$, $\eta$, $\dEdx$\} triplets 
is randomly generated according to the following procedure:
the momentum is generated according to a binned function based on selected tracks in  \textit{CR1};   
the pseudorapidity is generated according to the  $\eta$-binned (where $\eta$ depends on $p$) functions based on tracks in \textit{CR1}  and 
the ionisation is generated according to $\dEdx$-binned (where $\dEdx$ depends on $\eta$) functions based on all  tracks in \textit{CR2}.

For the $\sim$50000 random combinations in which $\dEdx$ is larger than the selection requirement, both in the stable and the metastable scenario, 
the particle mass  $m$ is obtained given the \{$\dEdx,p$\} generated values, using the technique explained in Section~\ref{sec:Strategy}.
The normalisation of the generated background distribution to the selected data is obtained by scaling the background distribution to the data
in the low-mass region of the mass distribution ($40<m<$~160~GeV) where a possible signal has  already been excluded~\cite{ATLAS:2013Rhad,Shimpei}.  
The normalisation is performed on the samples before the ionisation requirement, and its uncertainty 
 is dominated by the statistical uncertainty in the data. 
The complete procedure, from the key variables description, to the random generation, to the normalisation, is tested on signal-depleted regions.
These regions are the same as \textit{CR1} and  \textit{CR2} except for requiring tracks with  $100<p<150$~GeV instead of $p>150$~GeV.
Applied to these validation samples and the stable scenario, the procedure described above yields a predicted background of 48.9 $\pm$ 0.2 (stat.) $\pm$ 0.6 (norm.) 
events, while the number of events in the data sample is 49, for $m>160$~GeV.  
The same procedure applied to the metastable scenario yields a predicted background of 16.9 $\pm$ 0.1 (stat.) $\pm$ 0.4 (norm.) events, compared to 20 observed in the data, for $m>160$~GeV.


\section{Systematic uncertainties}\label{sec:Systematics}
Systematic uncertainties from several sources affecting the background estimate and the signal yield have been evaluated.
The uncertainties are quoted as the maximum deviation from the nominal expectation for the background or for the signal in the probed mass range. The actual systematic uncertainties are calculated and assigned per mass bin.

The uncertainties in the background estimation can be divided into three categories:
those related to the particular choices made for the binning, intervals, and fitting functions; 
those related to the different description of the key variables for control samples with different compositions than the search region sample; and those related to the stability as a function of pile-up.
The uncertainty on the background estimate due to each of these sources is evaluated by changing the description of the key variables, repeating the entire generation procedure, and comparing the resulting mass distribution with the nominal one. The uncertainties are estimated separately for the searches for stable and metastable particles. In each iteration, five million events are generated as explained in Sect.~\ref{sec:Background}. The resulting uncertainties are summarised in Table~\ref{tab:BkgSysSummary_stable}.

\begin{table}[!htbp]
\begin{center}
    \caption[]{Summary table of the  systematic uncertainties that affect the background estimations. All the sources are common to searches for stable and metastable particles, unless explicitly indicated. The uncertainties depend on the mass,  and the  maximum values are reported. In the limit calculations the actual value of the uncertainty for a given mass is used.\\
    \label{tab:BkgSysSummary_stable}}
 \vspace{0.3cm}
    \begin{tabular}{|l|r|}
       \hline
    {\bf Source of uncertainty:} &  {\boldmath {\bf[\%]}}\\
    \hline
    \hline
    Modification of: &   \\
    \hline
     - Binning in $p~$       &  $5$\\
     - Binning in $\eta$   &  $5$\\	
     - Momentum intervals in $\eta$ & $2$\\	
     - Binning in $\dEdx$   & $4$\\	
     - Analytical description of $\dEdx$ &   $2$\\	
    \hline
    Different fractions of: & \\
    \hline
     - Non-leptons in \textit{CR2} (metastable only)& $2$\\	
     - Tracks with more than 3 pixel clusters in \textit{CR2} & $1.5$\\	
     - Non-leptons in \textit{CR1} (stable only) & $4$\\	
     - $W$ decays in \textit{CR1} and \textit{CR2} & $3$\\	
    \hline
    Pile-up dependence & $4$\\	
    \hline
    \hline
    {\bf Total background systematic uncertainty:} & \\
    Stable particle search & $11$\ \\
    Metastable particle search & $10$\ \\
    \hline
    \end{tabular}
    \end{center}
\end{table}

The uncertainties on the signal yield are summarised in Table~\ref{tab:SysSummary_Rhad} and can be divided into three categories: 
those on the phenomenological modelling of the signal process with Monte Carlo generators;
those on the modelling of the detector efficiency or calibration; and
those affecting the overall signal yield.

The uncertainty on QCD radiation is evaluated for stable and metastable $R$-hadrons as the difference in efficiency between the \textsc{Madgraph} and 
\textsc{Pythia6} samples. This uncertainty is large for the benchmark channels in which the $\MET$ is dominated by the ISR contribution, such as the stable $R$-hadrons or the gluino  $R$-hadrons decaying into $g/q\bar{q}$ plus a heavy neutralino of mass $m(\tilde{\chi}^0_1)=m(\tilde{g})-100$~GeV.
The uncertainty on  QCD radiation is evaluated for metastable charginos using the same procedure as described in Ref.~\cite{Shimpei}, while for stable charginos the same uncertainty used for the stable $R$-hadrons is used, as the $\MET$ distributions are very similar. 
For the $R$-hadrons there is an uncertainty on how they interact with the detector material, and this is evaluated by comparing the efficiencies obtained from generating events according to the three scattering models described in Section~\ref{sec:R-hadrons}. Efficiencies for chargino events generated with and without the jet filter (see Section~\ref{sec:charginos_meta}) are compared, and their difference is accounted for as a systematic uncertainty.
A systematic uncertainty is assigned to the lifetime reweighting procedure, and is estimated as the discrepancy between the efficiencies obtained for the same reweighted lifetime starting from different  samples.

\begin{table}[!htbp]
    \caption[]{Summary table for the sources of systematic uncertainty considered for $R$-hadrons and charginos. The values are separately indicated for metastable and stable cases when these are different. The uncertainty depends on the mass and on the decay model, and the maximum negative and positive values are reported. In the limit calculations the actual value of the uncertainty for a given mass is used.    \label{tab:SysSummary_Rhad}}
  \vspace{0.3cm}
   \begin{tabular}{|l|cc|}
       \hline
    {\bf Source of uncertainty} & {\boldmath {\bf$-$[\%]}}& {\boldmath {\bf$+$[\%]}}\\
    \hline
    \hline
    {\bf $R$-hadron}&&\\
   QCD radiation modelling (stable) &$ -28$ & 28  \\
   QCD radiation modelling  (metastable) & $-12$ &12  \\
    Scattering models & $-9.9$&6.6 \\
    Lifetime reweighting (metastable)    & $-10$&10 \\
    \hline
    {\bf chargino}&&\\
    QCD radiation modelling (stable) & $-27$&27  \\
    QCD radiation modelling (metastable) & $-21$&30  \\
    Generator jet filter  & $-8.3$ &0\\
    Lifetime reweighting (metastable)    & $-10$&10 \\
  \hline
    Trigger efficiency modelling & $-4.5$&4.5 \\
    $\MET$ scale & $-3.8$&3.5 \\
    Pile-up   & $-1.7$&1.7 \\
    Ionisation parameterisation  & $-5.8$&0 \\
    Momentum parameterisation  (stable) & $-1.0$&1.0 \\
    Momentum parameterisation   (metastable) & $-2.0$&2.0 \\
    Track Efficiency parameterisation   & $-2.0$&2.0 \\    
    Electron identification    & $-1.0$&1.0 \\
    Muon identification (metastable only)    & $-1.0$&1.0 \\
    \hline   
     {\bf Total systematic uncertainty}&& \\
    \hline
      \emph{\textbf{Stable $R$-hadron}} & $-30$&29 \\
     \emph{\textbf{Metastable $R$-hadron}} & $-18$&15
 \\
    \hline
     \emph{\textbf{Stable chargino}} & $-30$&28 \\
     \emph{\textbf{Metastable chargino}} & $-24$&31 \\
    \hline
    {\bf Uncertainties on  signal yield} &  &\\
    Luminosity &  $-2.8$&2.8\\
    Cross section uncertainty ($R$-hadron)& $-56$&56 \\
    Cross section uncertainty (chargino)& $-8.5$&8.5 \\
       \hline
    \end{tabular}
\end{table}

The systematic uncertainties on the detector modelling are dominated by the trigger efficiency modelling, by the $\MET$ scale, and by the parameterisation of the ionisation. Systematic uncertainties related to the trigger are evaluated by varying the threshold and resolution parameters of the calorimetric $\MET$ trigger efficiency modelling curve and then looking at the efficiency difference between data and MC simulated $Z\to\mu^+\mu^-$ events. Systematic uncertainties of the $\MET$ measurement are evaluated with the methods described in Refs.~\cite{ATLAS:etmiss,ATLAS-CONF-2013-082} and are propagated to the uncertainty of the efficiency. Since the pile-up distribution is different in data and MC simulation, the simulated samples are reweighted to match the data, and a systematic uncertainty is calculated by varying the weighting factors. The systematic uncertainty on the pixel ionisation is evaluated by  comparing the ionisation of simulated and real tracks. These tracks are compatible with MIPs, selected with the same requirements as those used for the search. 
Other smaller uncertainties are due to momentum~\cite{ATLAS:2013Rhad} and track efficiency~\cite{ATLAS:tracking} parameterisation, and electron~\cite{Aad:2011dm} and muon~\cite{ATLAS_muons_2014} identification.

The uncertainty on the signal cross section is calculated as described in Section~\ref{sec:Simulation}. The uncertainty ranges from 15\% (at 100~GeV) to 56\% (at 1700~GeV) for $R$-hadrons, and is $\approx$8.5\% for the charginos, slightly dependent on the mass.


\section{Results}\label{sec:results}
\begin{figure}[htbp]
\resizebox{0.50\textwidth}{!}{%
   \includegraphics{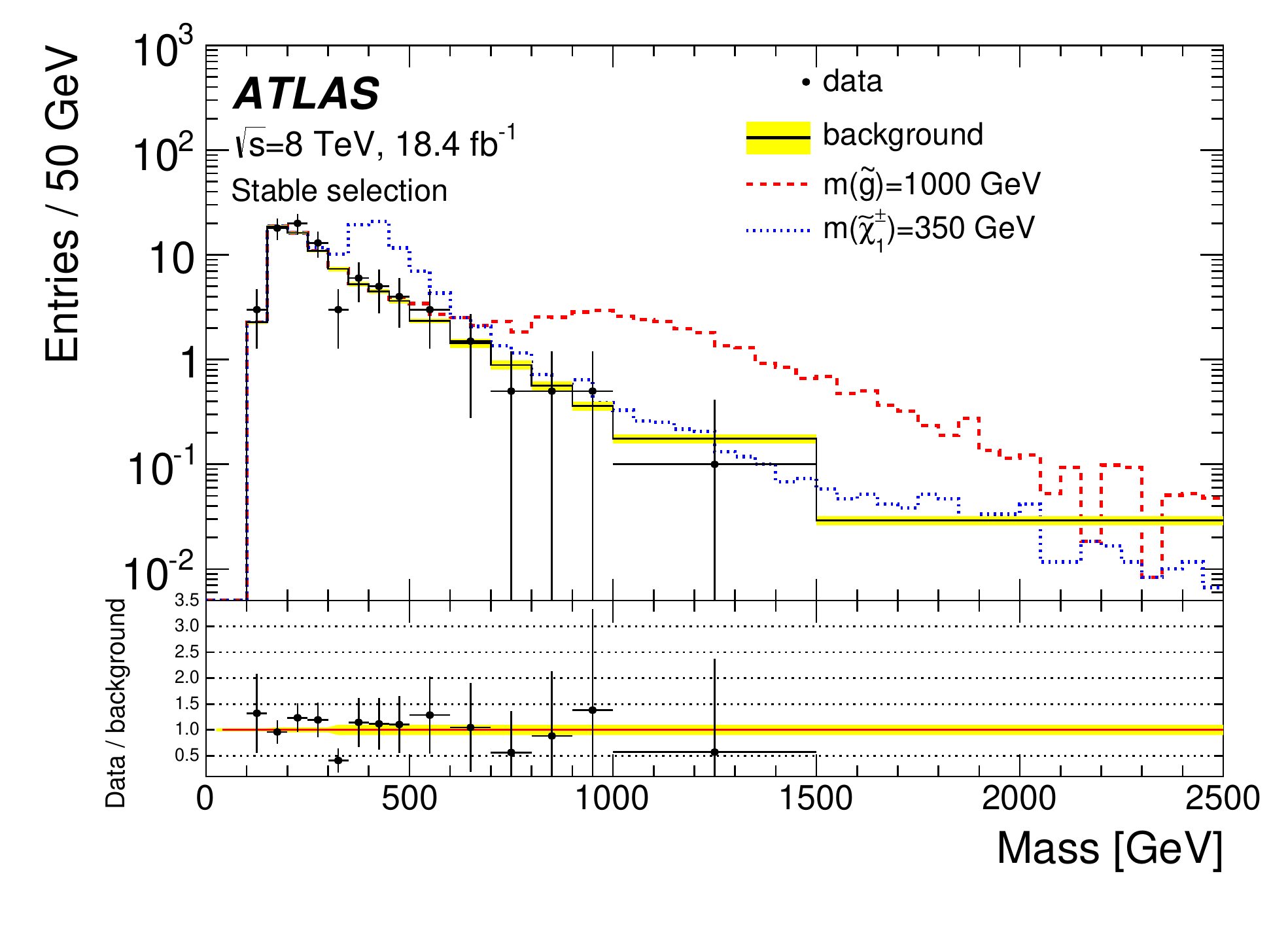}
 }
 \resizebox{0.50\textwidth}{!}{%
   \includegraphics{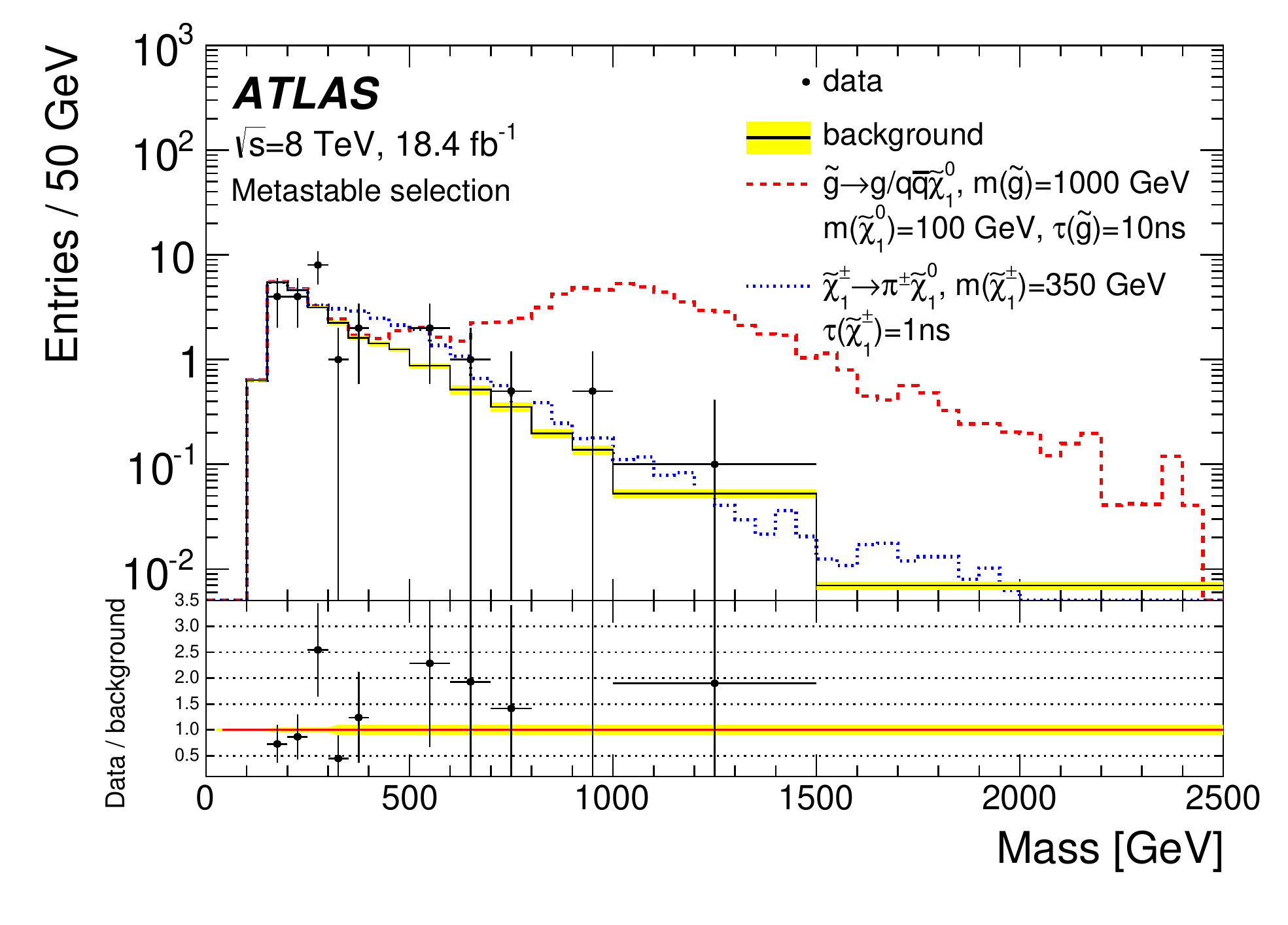}
}
  \caption{Distribution of the mass of selected candidates, derived from the specific ionisation loss, for data, background, and examples of gluino $R$-hadron and chargino signals, for searches for stable (top) and metastable (bottom) particles. The expected background is shown with its total uncertainty (sum in quadrature of statistical, normalisation and systematic errors). The signal distributions are stacked on the expected background, and a narrower binning is used for them to allow the signal shape to be seen more clearly. The number of signal events is that expected according to the theoretical cross sections. For both distributions, the  bin-per-bin ratio of data to expected background is also shown.}
  \label{fig:finalMass}
\end{figure}

The mass distribution is shown in Figure~\ref{fig:finalMass} for the selection of stable and metastable particles. The data, 85 events for the stable selection and 28 events for the metastable selection, are compared to the background estimate. In addition, mass distributions for examples of gluino and chargino signals are shown. 

No evidence of a signal above the expected background is observed.
The largest deviation has a local p-value of 4.3\% and occurs at a mass of 700~GeV for metastable gluino $R$-hadrons with lifetime $\tau =10$~ns.

Upper limits at 95\% confidence level (CL)  on $R$-hadron or chargino production cross sections (and, therefore, given a model cross section, a lower limit on their masses) are extracted using the CL$_S$ method ~\cite{Read:2002hq} by scanning the signal strength over a suitable range and using the profile likelihood ratio as a test statistic. 
In the procedure for setting the limit, the systematic uncertainties on the signal and background yields, 
as evaluated in  Section~\ref{sec:Systematics}, are treated as  Gaussian-distributed nuisance parameters. The statistical uncertainty on the background distribution also takes into account the uncertainty due to the normalisation.
Signal, background and data events are counted in a mass window of $\pm1.4\sigma$ around the signal peak, where the signal peak and the width are estimated by a Gaussian fit to the mass distribution in simulated signal MC samples.
Lower limits on the mass are derived by comparing the measured cross-section limits to the lower edge of the $\pm1\sigma$ band around the theoretically predicted cross section for each process.
The resulting lower limits set on the mass of the stable and metastable particles are summarised in Table~\ref{tab:massexcl} for the different searches.

\begin{table}[htdp]
\caption{The 95\% CL lower limit on the relevant LLP mass for the different models considered. Other relevant parameters (decay mode, neutralino mass, lifetime if metastable) are also shown. } \label{tab:massexcl}
 \vspace{0.3cm}
   \begin{tabular}{|rrrr|c| }
\hline
\bf{Particle}         & {\bf Decay}&{\bf{\boldmath $m(\tilde{\chi}^0_1$)\,[GeV] }}&{\bf {\boldmath$\tau$\,[ns]}} &{\boldmath {\bf $m>$\,[GeV]}}\\
\hline
$\tilde{g}~R$-hadron     & stable            & --                                      & --                & 1115 \\
$\tilde{b}~R$-hadron  & stable            & --                                      & --                & 751 \\
$\tilde{t}~R$-hadron        & stable            & --                                      & --                & 766 \\
\hline
chargino                 & stable            & --                                      & --                & 534 \\
\hline
 $\tilde{g}~R$-hadron     & $g/q\bar{q}$  & 100                                  & 10               & 1185 \\
 $\tilde{g}~R$-hadron     & $g/q\bar{q}$  & $m(\tilde{g})-100$           & 10               & 1099 \\
 $\tilde{g}~R$-hadron     & $t\bar{t}$       & 100                                  & 10               & 1182 \\
 $\tilde{g}~R$-hadron     & $t\bar{t}$       & $m(\tilde{g})-480$           & 10               & 1157 \\
 $\tilde{g}~R$-hadron     & $g/q\bar{q}$  & 100                                  & 1.0                & 869 \\
 $\tilde{g}~R$-hadron     & $g/q\bar{q}$  & $m(\tilde{g})-100$           & 1.0                & 821 \\
 $\tilde{g}~R$-hadron     & $t\bar{t}$       & 100                                  & 1.0                & 836 \\
 $\tilde{g}~R$-hadron     & $t\bar{t}$       & $m(\tilde{g})-480$           & 1.0                & 836\\
\hline
chargino      & $\tilde{\chi}^0_1 + \pi^\pm$    & $m(\tilde{\chi}^\pm_1)-0.14$           & 1.0                & 239 \\
chargino      & $\tilde{\chi}^0_1 + \pi^\pm$    & $m(\tilde{\chi}^\pm_1)-0.14$           & 15                & 482 \\
\hline
  \end{tabular}
\end{table}%

\begin{figure}[htbp]
\resizebox{0.50\textwidth}{!}{%
  \includegraphics{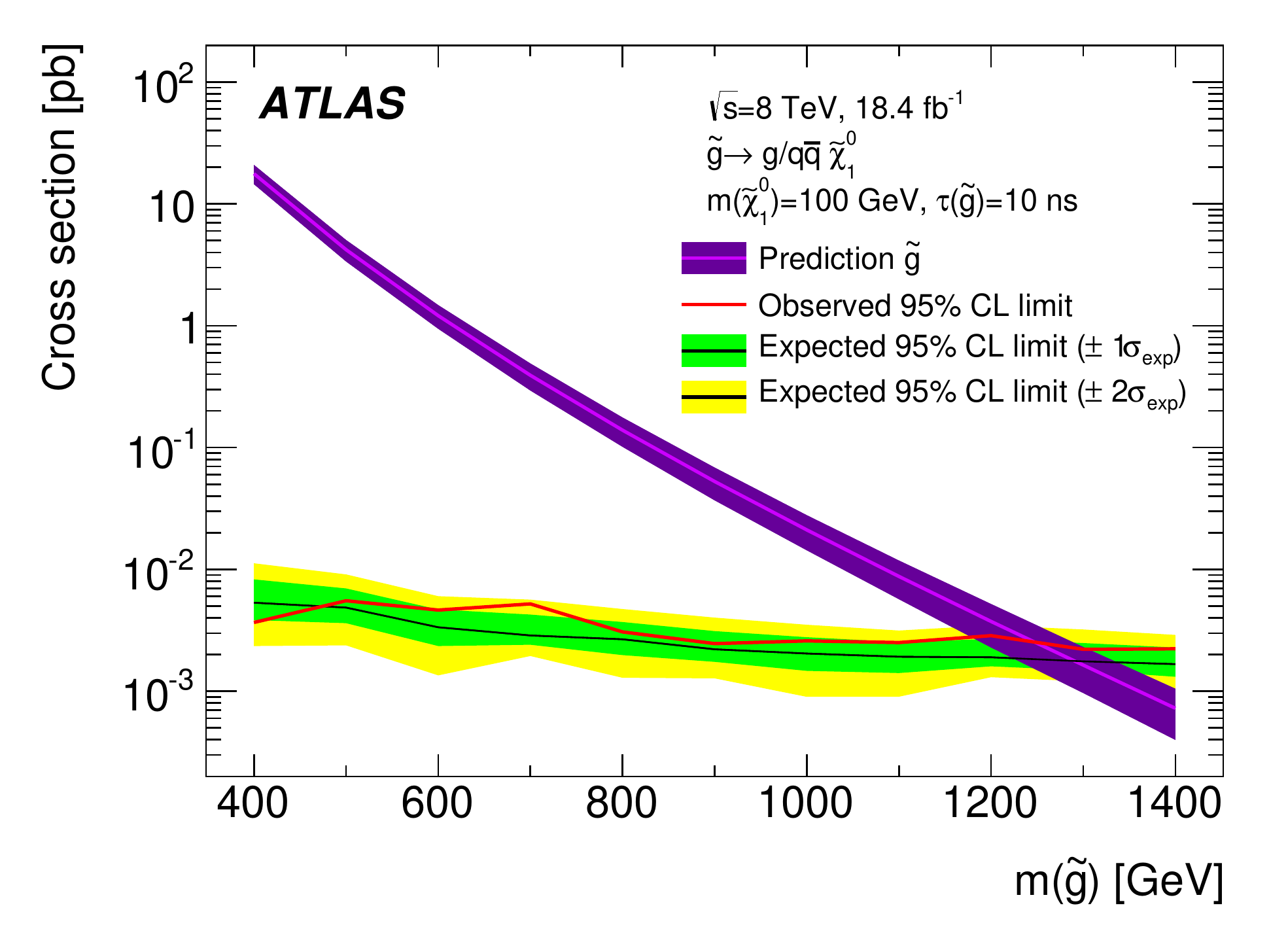}  
  } 
\resizebox{0.50\textwidth}{!}{%
   \includegraphics{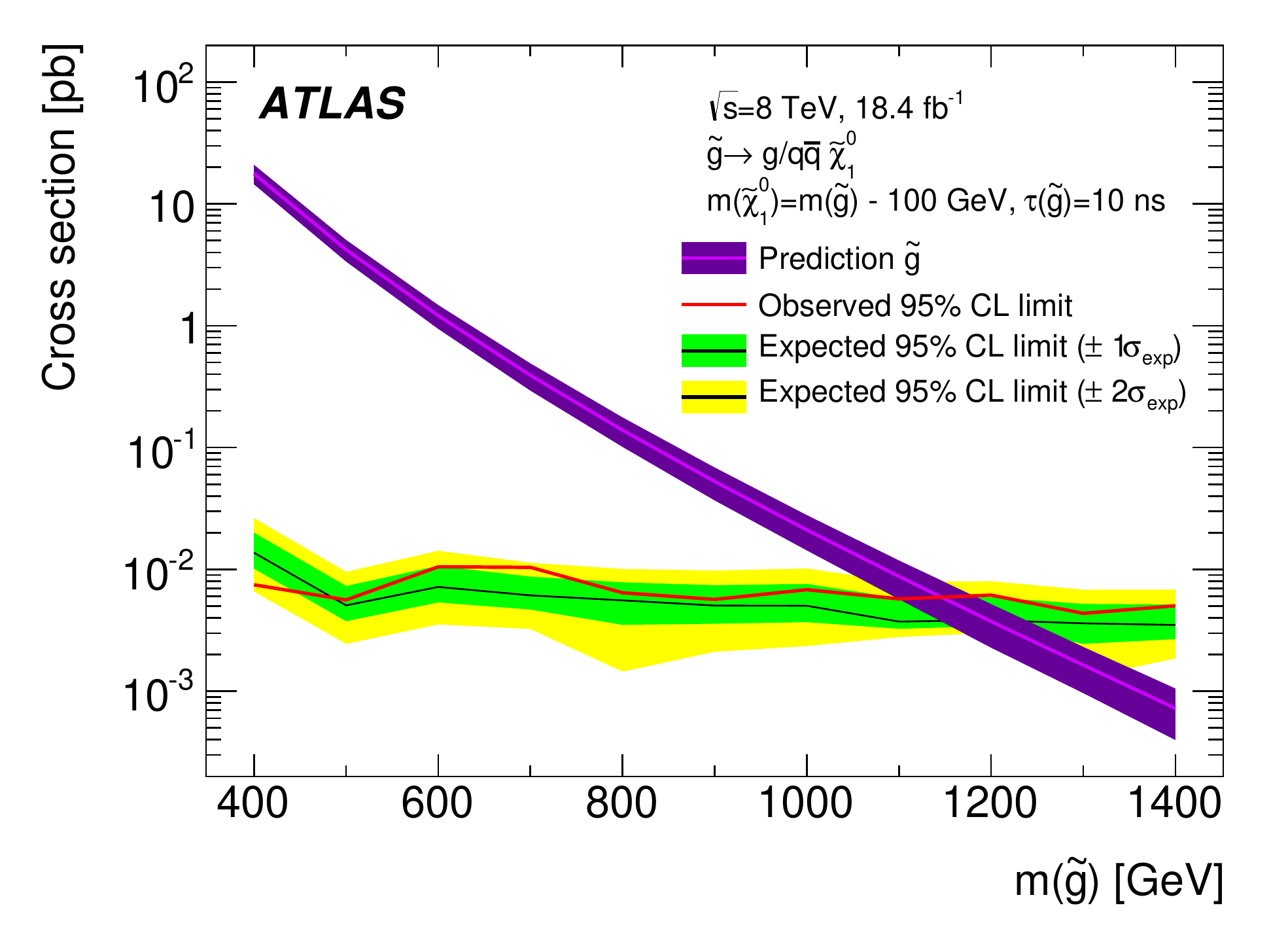}
 }
  \caption{Upper limits on the production cross section as a function of mass for metastable gluino $R$-hadrons, with lifetime $\tau =10$~ns, decaying into $g/q\bar{q}$ plus a light neutralino of mass $m(\tilde{\chi}^0_1)=100$~GeV (top) or a heavy neutralino of mass $m(\tilde{\chi}^0_1)=m(\tilde{g}) -100$~GeV (bottom). Theoretical values for the cross section are shown with their uncertainty. The expected upper limit in the background-only case is shown as a solid black line, with its $\pm 1\sigma$ and $\pm 2\sigma$ bands, green and yellow, respectively. The observed 95\%  CL upper limit is shown as a solid red line.}
  \label{fig:limmass_gluino10ns}
\end{figure}

\begin{figure}[htbp]
\resizebox{0.50\textwidth}{!}{%
  \includegraphics{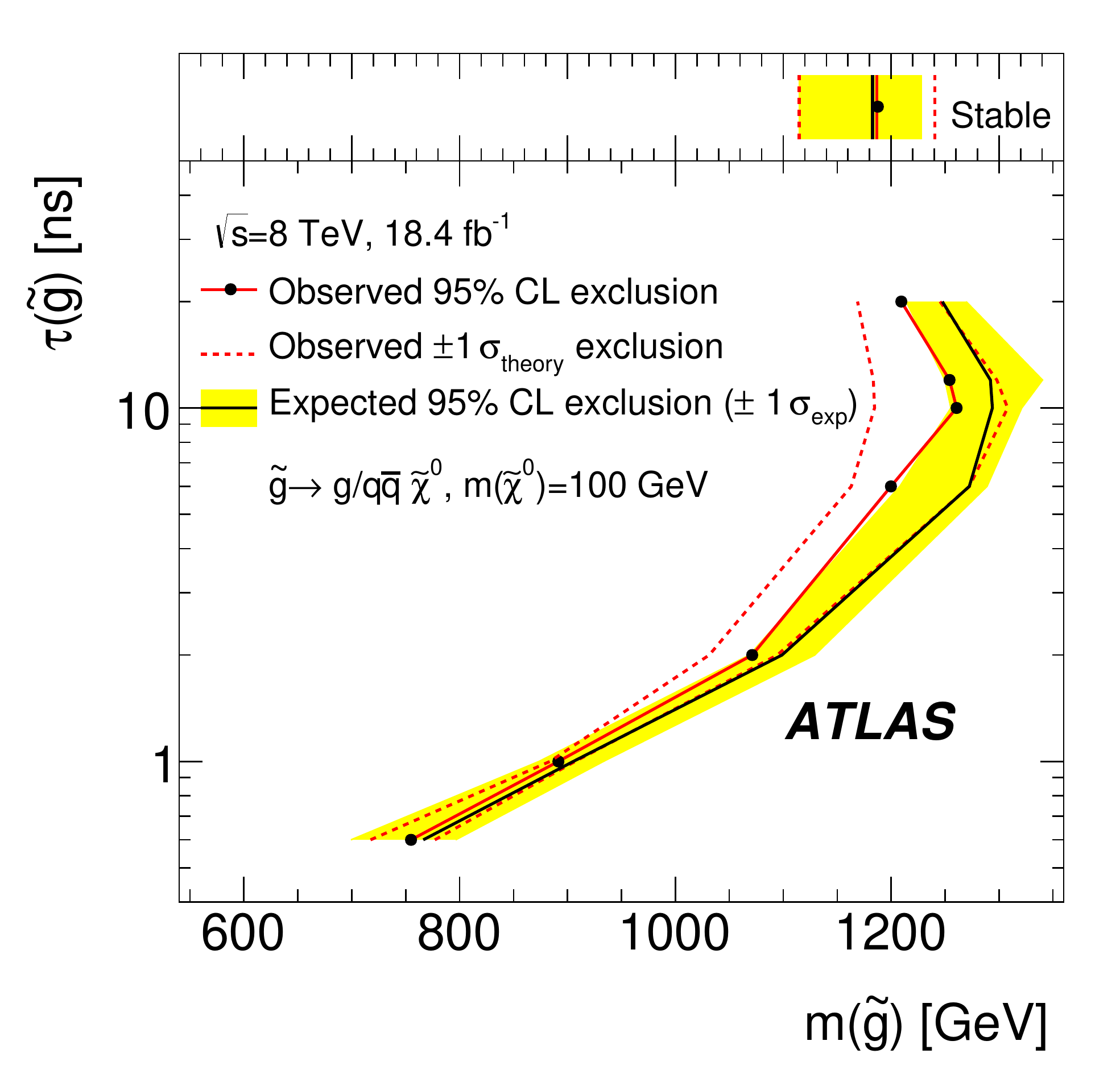}  
  }
 \resizebox{0.50\textwidth}{!}{%
 \includegraphics{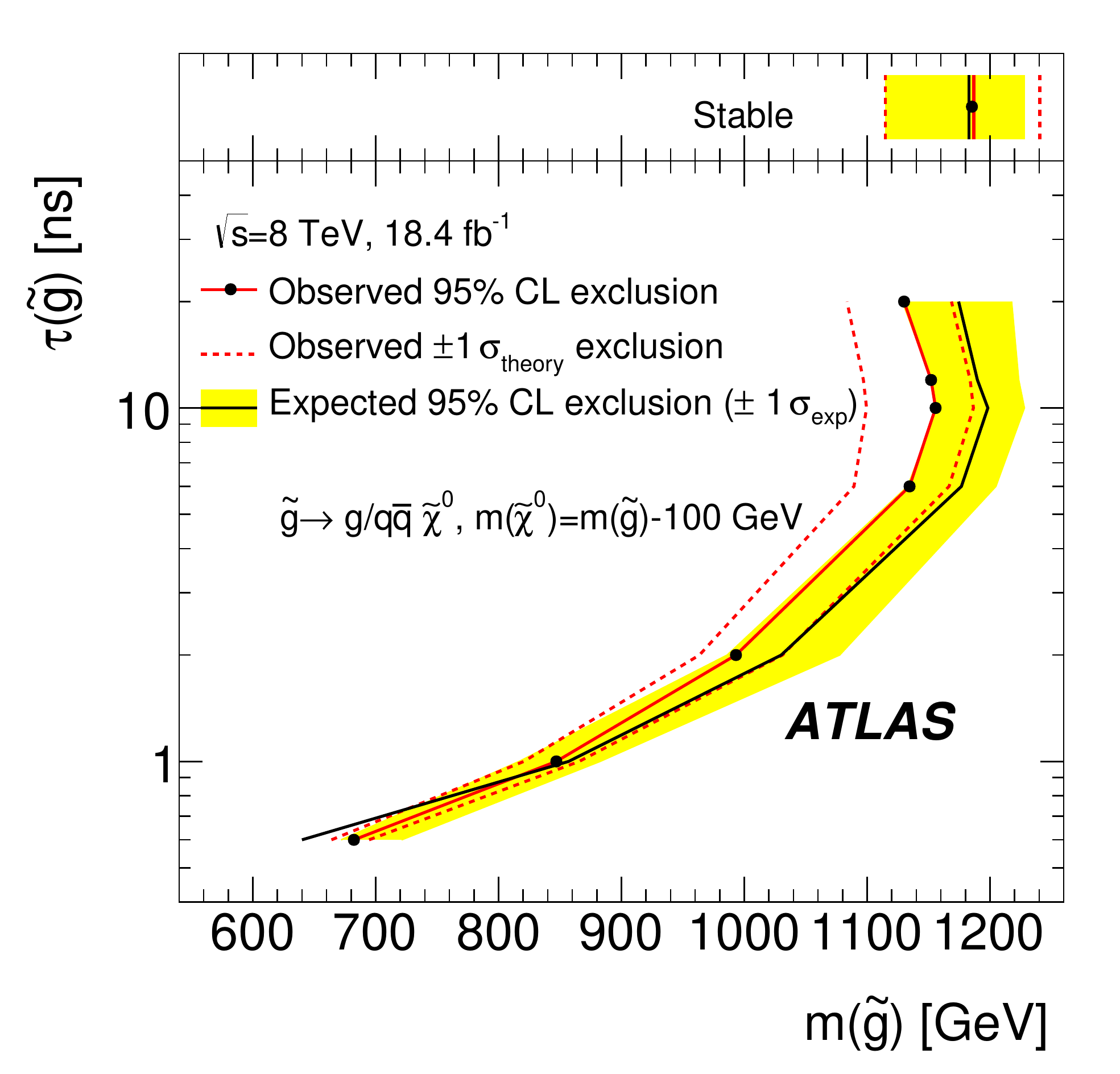}
}
  \caption{The excluded range of lifetimes as a function of gluino mass for gluino $R$-hadrons decaying into $g/q\bar{q}$ plus a light neutralino of mass $m(\tilde{\chi}^0_1)=100$~GeV (top) or a heavy neutralino of mass $m(\tilde{\chi}^0_1)=m(\tilde{g}) -100$~GeV (bottom). The expected exclusion, with its experimental $\pm1\sigma$ band, is given with respect to the nominal theoretical cross section.}
  \label{fig:lifemass_gluino}
\end{figure}

\begin{figure}[htbp]
\resizebox{0.50\textwidth}{!}{%
  \includegraphics{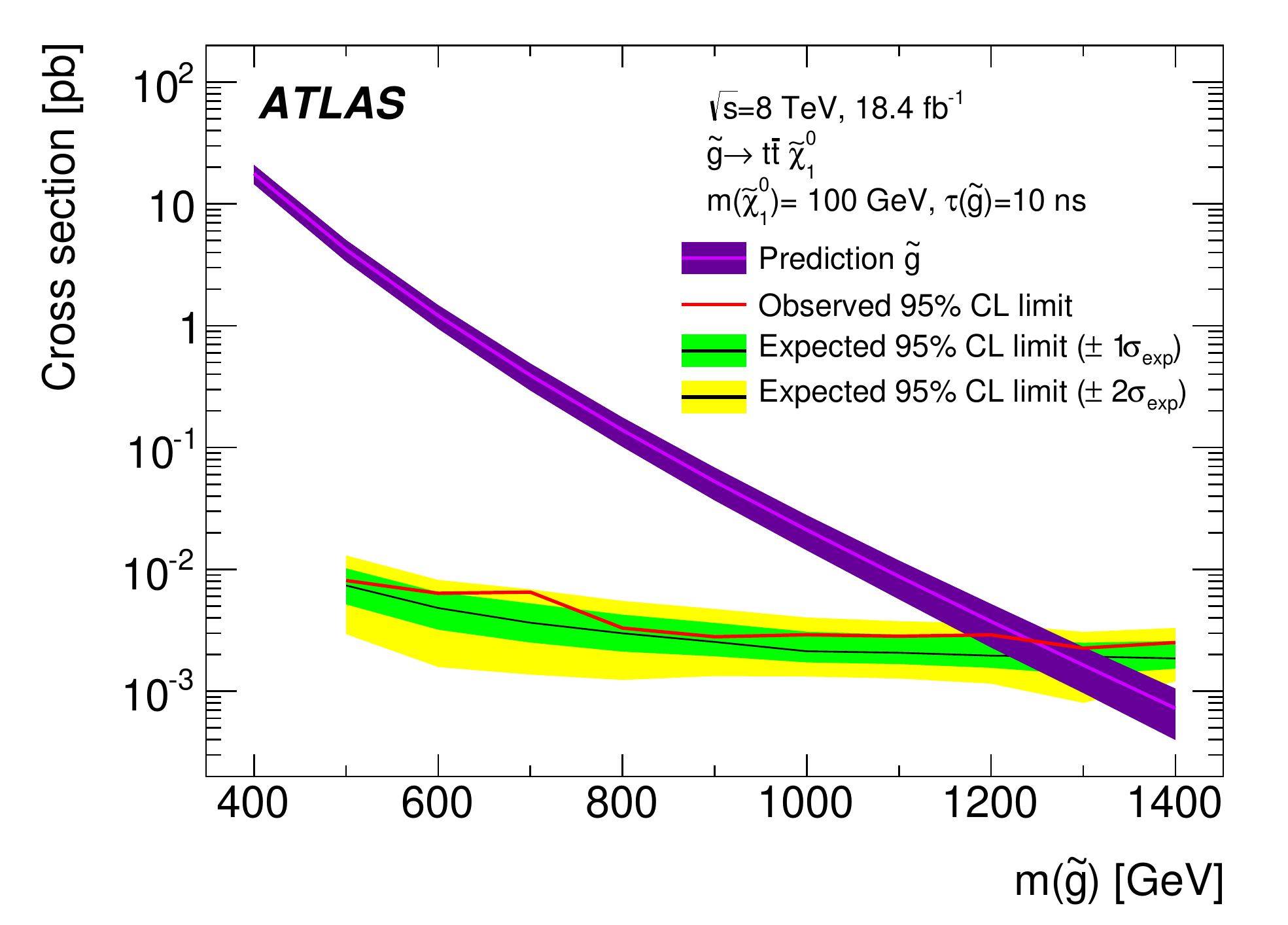}  
  } 
\resizebox{0.50\textwidth}{!}{%
   \includegraphics{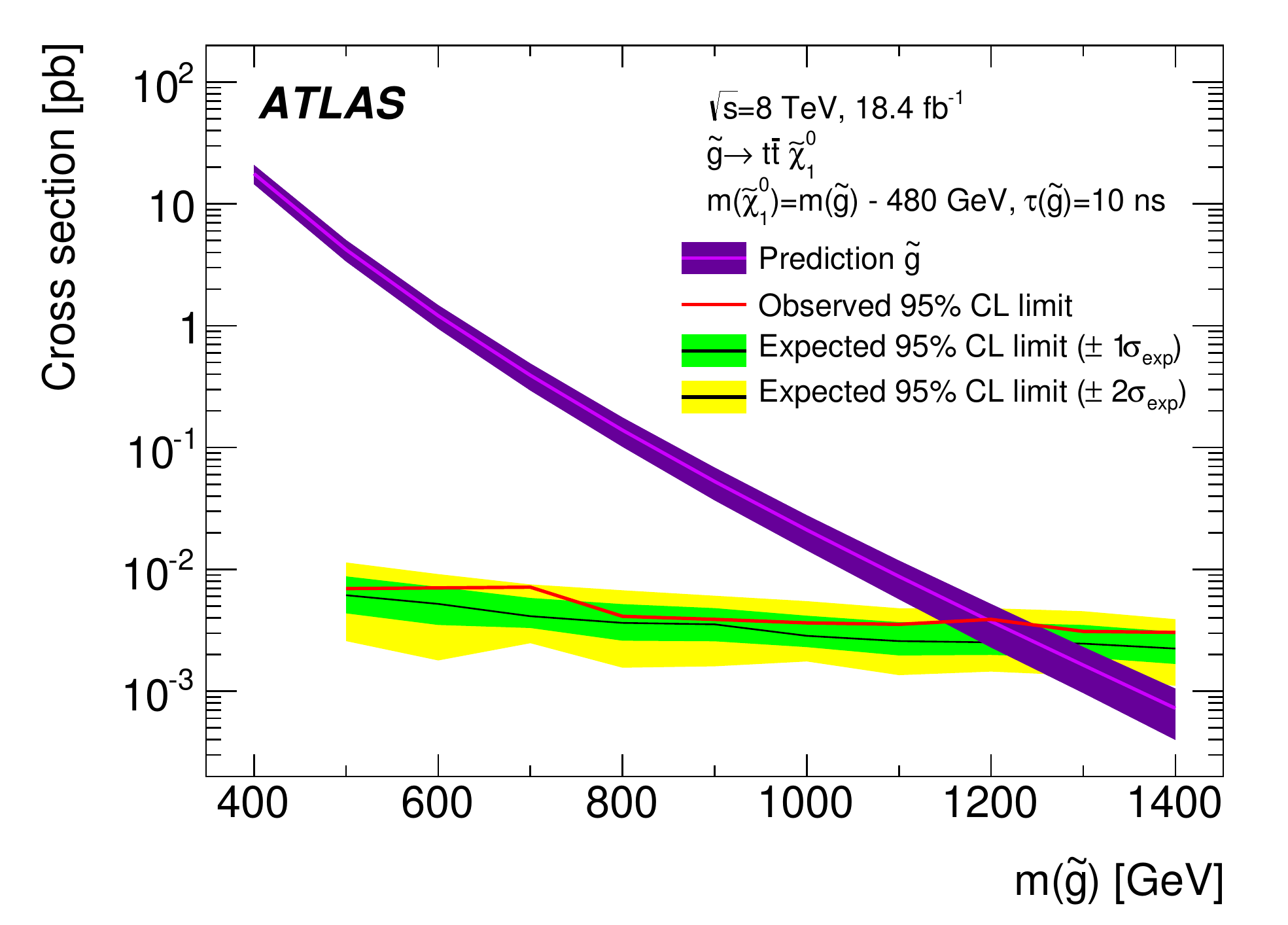}
 }
  \caption{Upper limits on the production cross section as a function of mass for metastable gluino $R$-hadrons, with lifetime $\tau =10$~ns, decaying into $t\bar{t}$ plus a light neutralino of mass $m(\tilde{\chi}^0_1)=100$~GeV (top) or a heavy neutralino of mass $m(\tilde{\chi}^0_1)=m(\tilde{g}) -480$~GeV (bottom). Theoretical values for the cross section are shown with their uncertainty. The expected upper limit in the background-only case is shown as a solid black line, with its $\pm 1\sigma$ and $\pm 2\sigma$ bands, green and yellow, respectively. The observed 95\% CL  upper limit is shown as a solid red line.}
  \label{fig:limmass_gluino10ns_tt}
\end{figure}

\begin{figure}[htbp]
\resizebox{0.50\textwidth}{!}{%
  \includegraphics{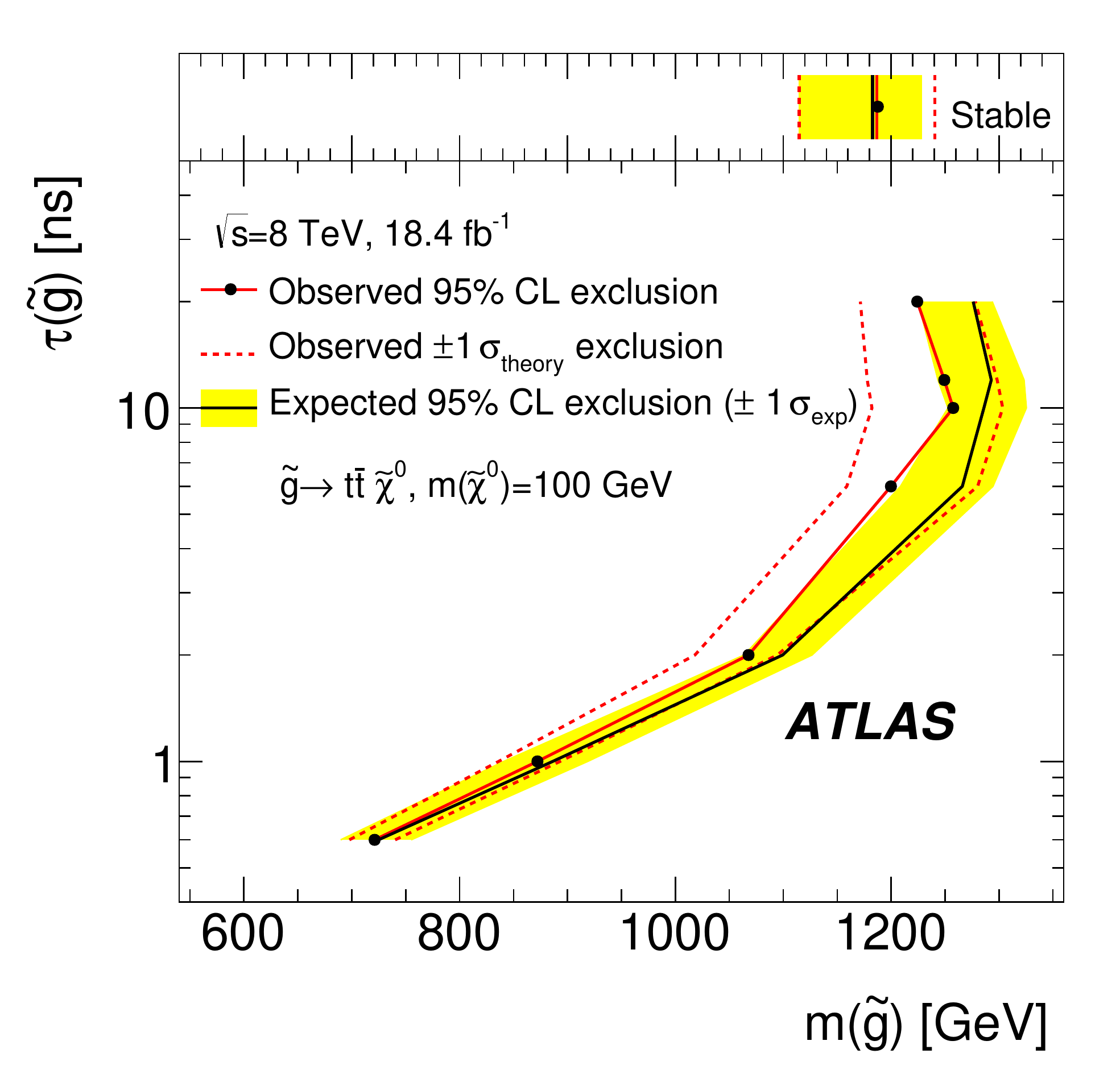} 
  }
 \resizebox{0.50\textwidth}{!}{%
 \includegraphics{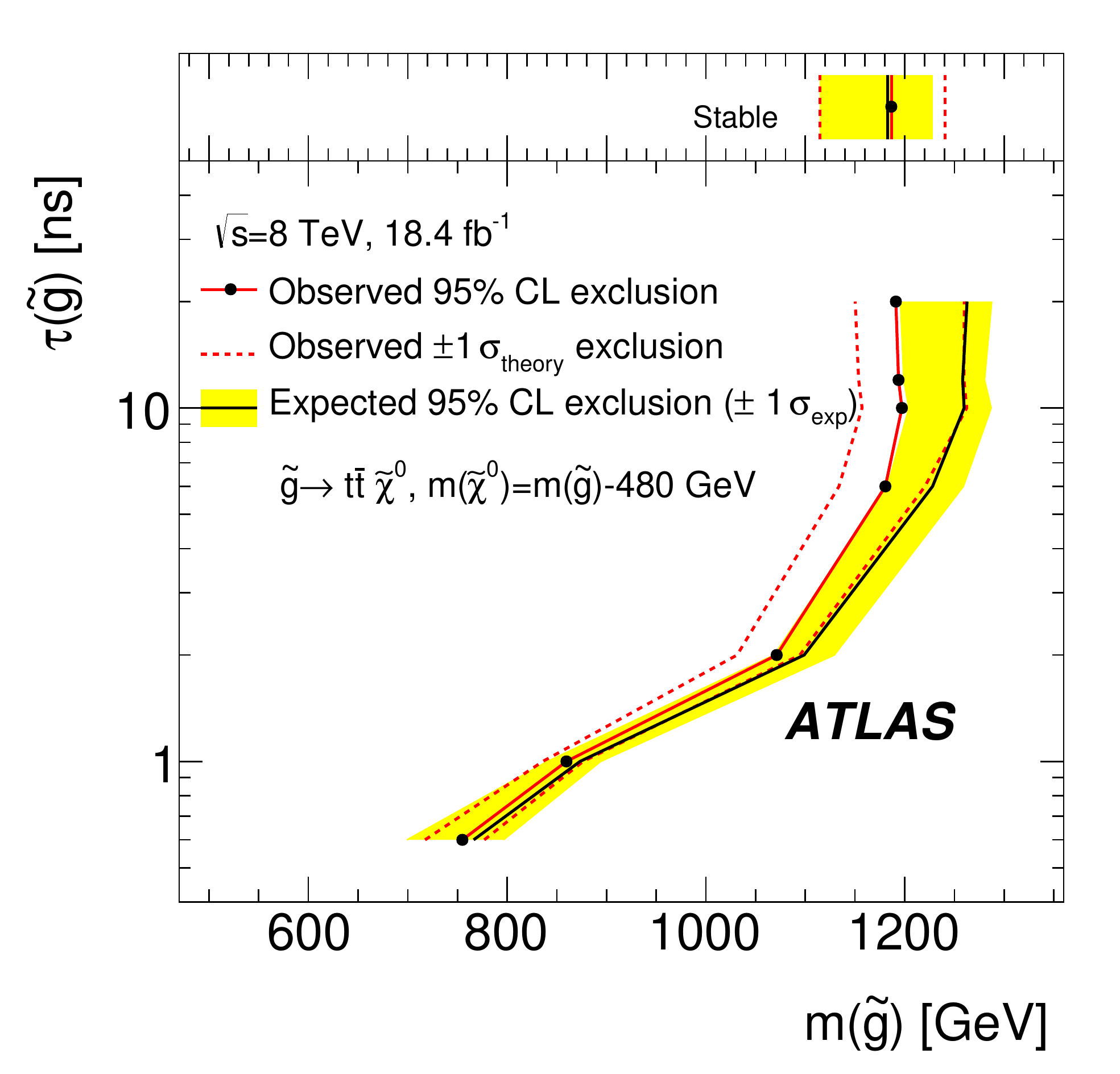}
}
  \caption{The excluded range of lifetimes as a function of gluino mass for gluino $R$-hadrons decaying into $t\bar{t}$ plus a light neutralino of mass $m(\tilde{\chi}^0_1)=100$~GeV (top) or a heavy neutralino of mass $m(\tilde{\chi}^0_1)=m(\tilde{g})-480$~GeV (bottom). The expected exclusion, with its experimental $\pm1\sigma$ band, is given with respect to the nominal theoretical cross section.}
  \label{fig:lifemass_gluino1}
\end{figure}

\begin{figure}[htbp]
\resizebox{0.50\textwidth}{!}{%
  \includegraphics{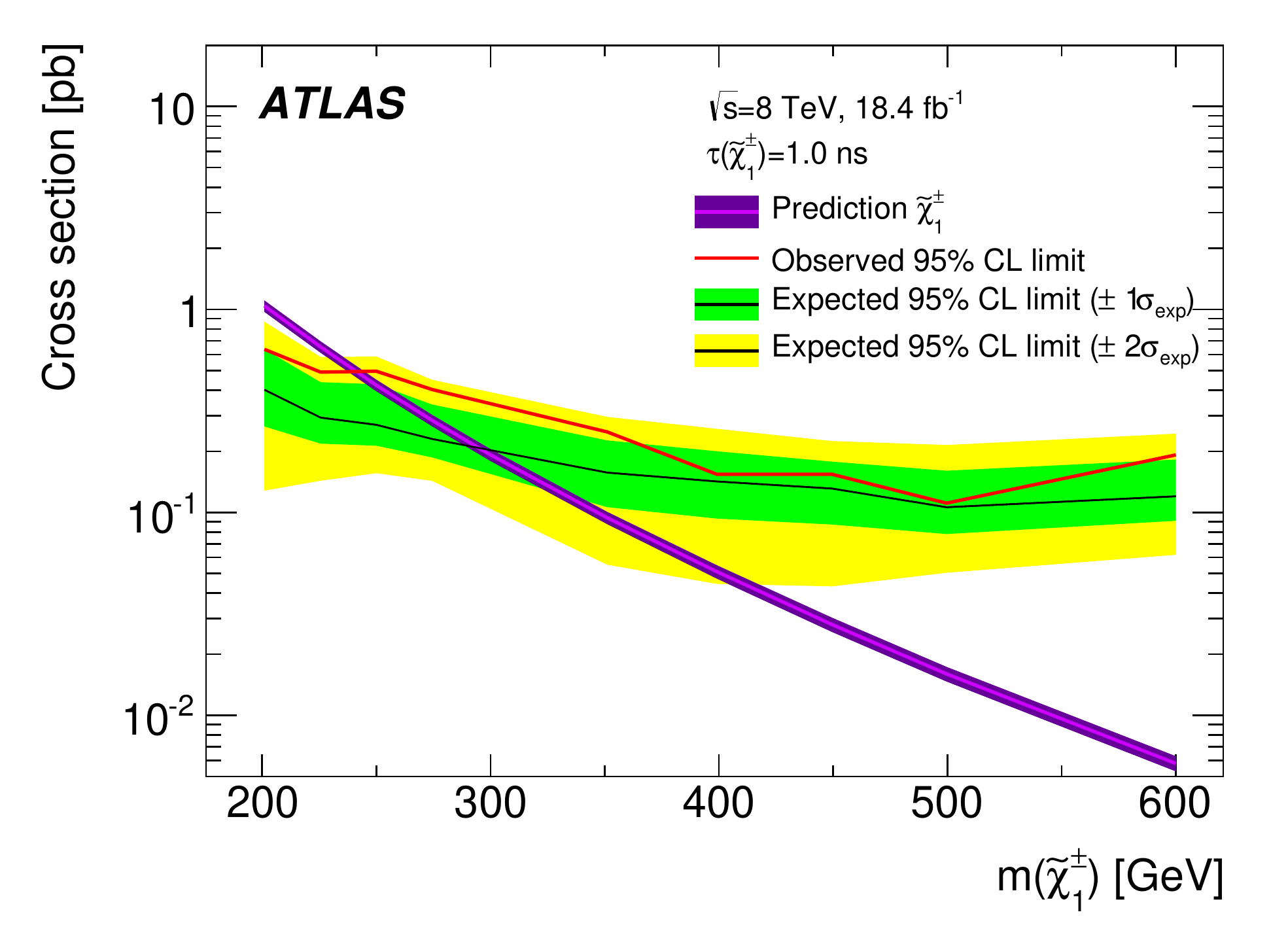}  
  } 
  \caption{Upper limits on the production cross section as a function of mass  for metastable charginos, with lifetime $\tau =1.0$~ns, decaying into $\tilde{\chi}^0_1 + \pi^\pm$. Theoretical values for the cross section are shown with their uncertainty. The expected upper limit in the background-only case is  shown as a solid black line, with its $\pm 1\sigma$ and $\pm 2\sigma$ bands, green and yellow, respectively. The observed 95\% CL upper limit is shown as a solid red line.}
  \label{fig:limmass_chargino1ns}
\end{figure}

\begin{figure}[htbp]
\resizebox{0.50\textwidth}{!}{%
  \includegraphics{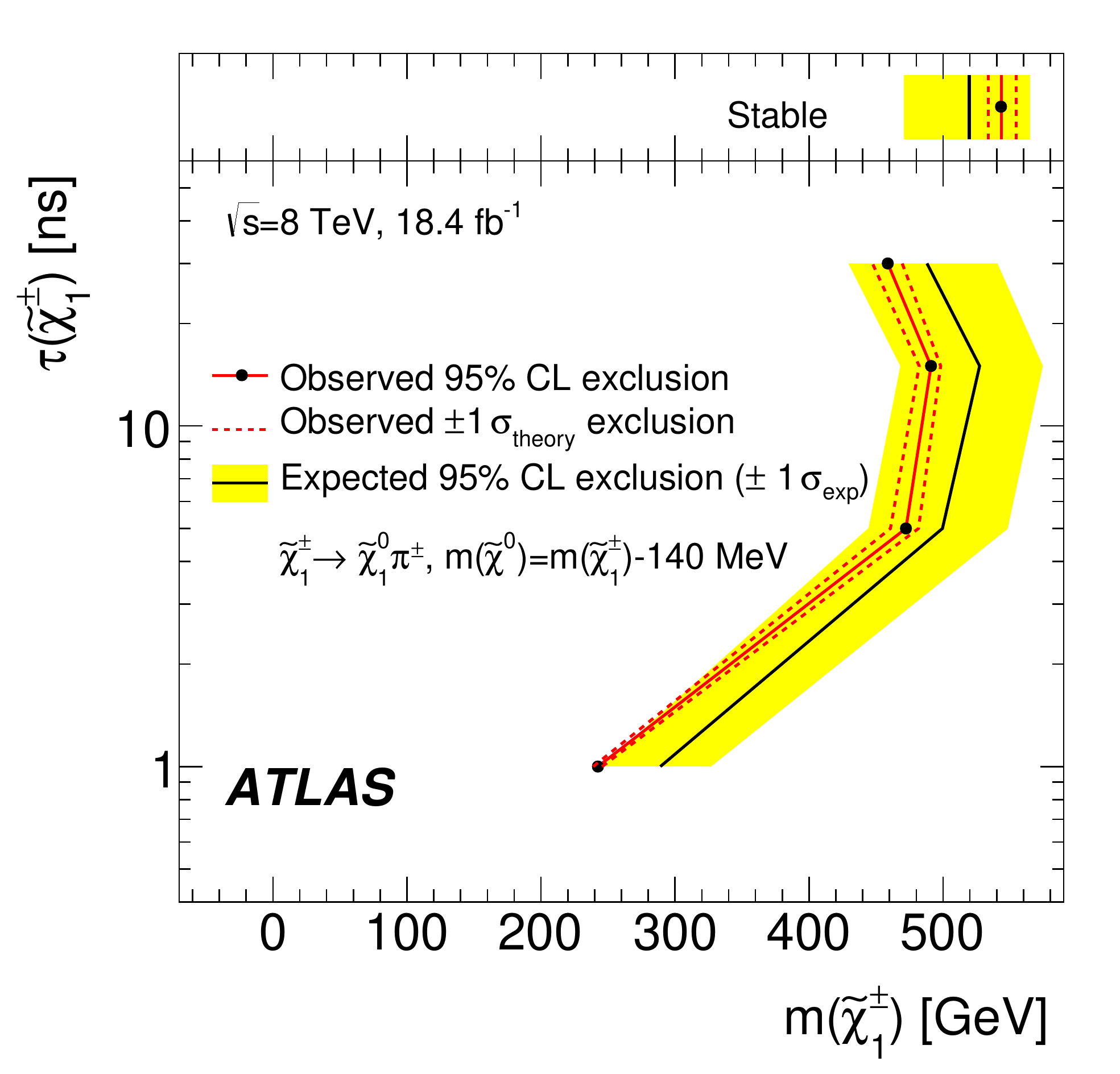}  
  } 
  \caption{The excluded range of lifetimes as a function of chargino mass for charginos decaying to $\tilde{\chi}^0_1 + \pi^\pm$. The expected exclusion, with its experimental $\pm1\sigma$ band, is given with respect to the nominal theoretical cross section.}
  \label{fig:lifemass_chargino}
\end{figure}

Figure~\ref{fig:limmass_gluino10ns} shows the upper limits on the production cross section for the gluino $R$-hadron with a lifetime of 10~ns decaying into $g/q\bar{q}$ plus a light neutralino of mass $m(\tilde{\chi}^0_1)=100$~GeV, or a heavy neutralino of mass $m(\tilde{\chi}^0_1)=m(\tilde{g})-100$~GeV.
Figure~\ref{fig:lifemass_gluino} shows the excluded range of lifetimes for the same $R$-hadron decays versus the particle mass.
Figure~\ref{fig:limmass_gluino10ns_tt} shows the upper limits on the production cross section for the gluino $R$-hadron with a lifetime of 10~ns decaying into $t\bar{t}$ plus a light neutralino of mass $m(\tilde{\chi}^0_1)=100$~GeV, or a heavy neutralino of mass $m(\tilde{\chi}^0_1)=m(\tilde{g})-480$~GeV.
Figure~\ref{fig:lifemass_gluino1} shows the excluded range of lifetimes for the same $R$-hadron decays as a function of the particle mass. 
As shown in Figures~\ref{fig:lifemass_gluino} and \ref{fig:lifemass_gluino1} the sensitivity of the measurement for $R$-hadrons is maximal for lifetimes around 10~ns.
Figure~\ref{fig:limmass_chargino1ns} shows the upper limits on the production cross section for charginos with lifetime of 1~ns decaying into $\tilde{\chi}^0_1 + \pi^\pm$  and Figure~\ref{fig:lifemass_chargino} shows the excluded range of lifetimes as a function of the particle mass for the same chargino decay.

\clearpage
\section{Summary}
A search has been performed for stable and metastable non-relativistic long-lived charged particles identified through their anomalous specific ionisation energy loss in the ATLAS pixel detector. The search uses $\lumi$~fb$^{-1}$ of $pp$ collision data at $\sqrt{s} = 8$~TeV collected by the ATLAS detector at the LHC.
In the scenario considered, stable charginos with masses smaller than 534~GeV are excluded at 95\% confidence level, and so are stable $R$-hadrons with masses smaller than 1115~GeV for gluinos, 751~GeV for bottom squarks and 766~GeV for top squarks. 
For metastable particles the maximum sensitivity  is reached at 10~ns lifetime for $R$-hadrons, where masses below 1185~GeV are excluded, and at 15~ns lifetime for charginos, where masses up to 482~GeV are excluded. Compared to previous searches this search provides the best sensitivity for gluinos with lifetimes between 3 and 20ns.


\section{Acknowledgements}

We thank CERN for the very successful operation of the LHC, as well as the
support staff from our institutions without whom ATLAS could not be
operated efficiently.

We acknowledge the support of ANPCyT, Argentina; YerPhI, Armenia; ARC,
Australia; BMWFW and FWF, Austria; ANAS, Azerbaijan; SSTC, Belarus; CNPq and FAPESP,
Brazil; NSERC, NRC and CFI, Canada; CERN; CONICYT, Chile; CAS, MOST and NSFC,
China; COLCIENCIAS, Colombia; MSMT CR, MPO CR and VSC CR, Czech Republic;
DNRF, DNSRC and Lundbeck Foundation, Denmark; EPLANET, ERC and NSRF, European Union;
IN2P3-CNRS, CEA-DSM/IRFU, France; GNSF, Georgia; BMBF, DFG, HGF, MPG and AvH
Foundation, Germany; GSRT and NSRF, Greece; RGC, Hong Kong SAR, China; ISF, MINERVA, GIF, I-CORE and Benoziyo Center, Israel; INFN, Italy; MEXT and JSPS, Japan; CNRST, Morocco; FOM and NWO, Netherlands; BRF and RCN, Norway; MNiSW and NCN, Poland; GRICES and FCT, Portugal; MNE/IFA, Romania; MES of Russia and ROSATOM, Russian Federation; JINR; MSTD,
Serbia; MSSR, Slovakia; ARRS and MIZ\v{S}, Slovenia; DST/NRF, South Africa;
MINECO, Spain; SRC and Wallenberg Foundation, Sweden; SER, SNSF and Cantons of
Bern and Geneva, Switzerland; NSC, Taiwan; TAEK, Turkey; STFC, the Royal
Society and Leverhulme Trust, United Kingdom; DOE and NSF, United States of
America.

The crucial computing support from all WLCG partners is acknowledged
gratefully, in particular from CERN and the ATLAS Tier-1 facilities at
TRIUMF (Canada), NDGF (Denmark, Norway, Sweden), CC-IN2P3 (France),
KIT/GridKA (Germany), INFN-CNAF (Italy), NL-T1 (Netherlands), PIC (Spain),
ASGC (Taiwan), RAL (UK) and BNL (USA) and in the Tier-2 facilities
worldwide.

%
\bibliographystyle{atlasBibStyleWoTitle}
\bibliography{EPJC_metastable_pixel}

\clearpage 
\onecolumn
\begin{flushleft}
{\Large The ATLAS Collaboration}

\bigskip

G.~Aad$^{\rm 85}$,
B.~Abbott$^{\rm 113}$,
J.~Abdallah$^{\rm 151}$,
O.~Abdinov$^{\rm 11}$,
R.~Aben$^{\rm 107}$,
M.~Abolins$^{\rm 90}$,
O.S.~AbouZeid$^{\rm 158}$,
H.~Abramowicz$^{\rm 153}$,
H.~Abreu$^{\rm 152}$,
R.~Abreu$^{\rm 30}$
% ...
\bigskip
\\
% ...
$^{11}$ Institute of Physics, Azerbaijan Academy of Sciences, Baku, Azerbaijan\\
% ...
$^{30}$ CERN, Geneva, Switzerland\\
% ...
$^{85}$ CPPM, Aix-Marseille Universit{\'e} and CNRS/IN2P3, Marseille, France\\
% ...
$^{90}$ Department of Physics and Astronomy, Michigan State University, East Lansing MI, United States of America\\
% ...
$^{107}$ Nikhef National Institute for Subatomic Physics and University of Amsterdam, Amsterdam, Netherlands\\
% ...
$^{113}$ Homer L. Dodge Department of Physics and Astronomy, University of Oklahoma, Norman OK, United States of America\\
% ...
$^{151}$ Institute of Physics, Academia Sinica, Taipei, Taiwan\\
$^{152}$ Department of Physics, Technion: Israel Institute of Technology, Haifa, Israel\\
$^{153}$ Raymond and Beverly Sackler School of Physics and Astronomy, Tel Aviv University, Tel Aviv, Israel\\
% ...
$^{158}$ Department of Physics, University of Toronto, Toronto ON, Canada\\
% ...
\end{flushleft}


\end{document}